\documentclass{aa}
\usepackage{graphicx}
\usepackage{txfonts}
\usepackage{natbib}
\usepackage[usenames,dvipsnames]{color}
\usepackage{ulem}
\newcommand\irc{\object{IRC\,+10420 }}
\newcommand\ircend{\object{IRC\,+10420}}
\newcommand\brg{Br$\gamma\,$}
\newcommand\brge{Br$\gamma$}

\newcommand{\kms}{\ensuremath{{\rm km\,s^{-1}}}}                   % \kmsec
\newcommand{\msun}{\ensuremath{\mathrm{M}_{\odot}}}                  % solar masses: msun
\newcommand{\msunyr}{\ensuremath{\mathrm{M}_{\odot}{\rm yr}^{-1}}}   % msun/yr
\newcommand{\lsun}{\ensuremath{\mathrm{L}_{\odot}}}                  % solar luminosity
\newcommand{\geff}{\ensuremath{\mathrm{g}_{\rm eff}}}                % effective surface gravity
\newcommand{\logg}{\ensuremath{\log \mathrm{g}_{\rm eff}}}                     % log surface gravity
\newcommand{\lstar}{\ensuremath{\mathit{L}_{\star}}}                 % stellar luminosity
\newcommand{\mdot}{\ensuremath{\dot{M}}}                             % mass loss rate
\newcommand{\rstar}{\ensuremath{\mathit{R}_{\star}}}                 % stellar radius
\newcommand{\teff}{\ensuremath{\mathit{T}_{\rm eff}}}                % effectieve temperatuur
\newcommand{\vinf}{\ensuremath{v_{\infty}}}                          % maximale uistroomsnelheid
\begin{document}
%%%%%%%%%%%%%%%%%
   \title{Resolving the asymmetric inner wind region of the yellow hypergiant \irc
          with VLTI/AMBER in low and high spectral resolution mode\thanks{
          The low-spectral resolution data have been obtained as part of the Guaranteed Time Programme
          for VLTI/AMBER (program ID: 079.D-0356(B)), while the high-spectral resolution data were 
          obtained in the context of science verification observations (program ID: 60.A-9053(D)).}
}
\titlerunning{VLTI/AMBER observations of the yellow hypergiant \irc}
   \author{T.~Driebe\inst{1}
          \and
	J.~H.~Groh\inst{1}
          \and
	K.-H.~Hofmann\inst{1}
          \and
	K.~Ohnaka\inst{1}
          \and
	S.~Kraus\inst{1}
          \and
	F.~Millour\inst{1}
          \and
	K.~Murakawa\inst{1}
          \and
	D.~Schertl\inst{1}
          \and
    	G.~Weigelt\inst{1}
	  \and
	R.~Petrov\inst{2}
	  \and
        M.~Wittkowski\inst{3}
          \and
        C.A.~Hummel\inst{3}
	  \and
        J.B.~Le Bouquin\inst{4}
	  \and
        A.~Merand\inst{4}
          \and
        M.~Sch\"oller\inst{3}
          \and
        F.~Massi\inst{5}
          \and
        P.~Stee\inst{6}
          \and
        E.~Tatulli\inst{7}
          }

   \institute{Max-Planck-Institut f\"ur Radioastronomie, Auf dem H\"ugel 69, D-53121 Bonn, Germany, \\
              \email{driebe@mpifr-bonn.mpg.de}
         \and
   Laboratoire Universitaire d'Astrophysique de Nice, UMR 6525, Universit\'e de Nice/CNRS,
             06108 Nice Cedex 2, France
         \and
   European Southern Observatory, Karl-Schwarzschild-Str. 2, D-85748 Garching bei M\"unchen, Germany
         \and
   European Southern Observatory, Alonso de Cordova 3107, Vitacura, Casilla 19001, Santiago 19, Chile
         \and
   INAF-Osservatorio Astrofisico di Arcetri, Istituto Nazionale di Astrofisica, 
   Largo E.~Fermi 5, 50125 Firenze, Italy
	\and
	Observatoire de la C\^ote d'Azur/CNRS, UMR 6525 H. Fizeau, Univ. Nice Sophia Antipolis, 
        Avenue Copernic, 06130 Grasse, France	
	\and
   Laboratoire d'Astrophysique	de Grenoble, UMR 5571, Universit\'e Joseph Fourier/CNRS, 
   38041 Grenoble Cedex 9, France
}
   \date{Version: \today}

% \abstract{}{}{}{}{} 
% 5 {} token are mandatory
 
  \abstract
   {\irc is a massive evolved star belonging to the group of yellow hypergiants. Currently, this star
    is rapidly evolving through the Hertzprung-Russell diagram, crossing the so-called yellow void.
    \irc is suffering from intensive mass loss which led to the formation of an extended dust shell.
    Moreover, the dense stellar wind of \irc is subject to strong line emission.}
   {Our goal was to probe the photosphere and the innermost circumstellar environment of \ircend, to
    measure the size of its continuum- as well as the \brg line-emitting region on milliarcsecond
    scales, and to search for evidence of an asymmetric distribution of \ircend's dense, circumstellar gas.}
   {We obtained near-infrared long-baseline interferometry of \irc  
    with the AMBER instrument of ESO's Very Large Telescope Interferometer (VLTI).
    The measurements were carried out in May/June 2007 and May 2008 in low-spectral resolution mode in the $JHK$
    bands using three Auxillary Telescopes (ATs) at projected baselines ranging from 30 to 96~m, and in October 
    2008 in high-spectral resolution mode in the $K$ band around the \brg emission line 
    using three Unit Telescopes (UTs) with projected baselines between 54 and 129~m.
    The high-spectral resolution mode observations were analyzed by means of radiative transfer
    modeling using CMFGEN and the 2-D Busche \& Hillier codes.}
   {{For the first time, we have been able to absolutely calibrate the $H$- and $K$-band data and, thus, to determine 
    the angular size of IRC+10420's continuum- and Br$\gamma$ line-emitting regions. We found that both the low 
    resolution differential and closure phases are zero within the uncertainty limits across all three bands. 
    In the high-spectral resolution observations, the visibilities show a noticeable drop across the Br$\gamma$ line
    on all three baselines. We found differential phases up to -25\degr\, in the redshifted part of the Br$\gamma$ 
    line and a non-zero closure phase close to the line center. The calibrated visibilities were corrected for AMBER's 
    limited field-of-view to appropriately account for the flux contribution of \ircend's extended dust shell.
    From our low-spectral resolution AMBER data we derived FWHM Gaussian sizes of $1.05\pm 0.07$ and $0.98\pm 0.10$~mas 
    for \ircend's continuum-emitting region in the $H$ and $K$ bands, respectively.
    From the high-spectral resolution data, we obtained a FWHM Gaussian size of $1.014\pm0.010$~mas
    in the $K$-band continuum. The \brg-emitting region can be fitted with a geometric ring model with a diameter
    of $4.18^{+0.19}_{-0.09}~$mas, which is approximately 4 times the stellar size. 
    The geometric model also provides some evidence that the \brg line-emitting region is elongated towards a 
    position angle of 36\degr, well aligned with the symmetry axis of the outer reflection nebula. Assuming an unclumped 
    wind and a luminosity of $6\times10^5\lsun$, the spherical radiative transfer modeling with CMGFEN yields a current 
    mass-loss rate of $1.5 - 2.0\times 10^{-5}\msunyr$ based on the \brg equivalent width. However, the 
    spherical CMFGEN model poorly reproduces the observed line shape, blueshift, and extension, definitively 
    showing that the \irc outflow is asymmetric. Our 2-D radiative transfer modeling shows that the 
    blueshifted \brg emission and the shape of the visibility across the emission line can be explained with 
    an asymmetric bipolar outflow with a high density contrast from pole to equator (8--16), where the 
    redshifted light is substantially diminished.}}
   {}

\keywords{
Techniques: high angular resolution ---
Techniques: interferometric ---
Circumstellar matter ---
Stars: individual: IRC\,+10\,420 ---
Stars: mass--loss ---
Stars: supergiants ---
Stars: Wolf-Rayet
}
   \maketitle
%
%________________________________________________________________

%%%%%%%%%%%%%%%%%%%%%%%%%%%%%%%%%%%%%%%%%%%%%%%%%%%%%%%%%%%%%%%%%%%%%%%%%%%%%%%%%%%%%%%%%%%%%%%%%%%%%%%
\section{Introduction}\label{sect_intro}
%%%%%%%%%%%%%%%%%%%%%%%%%%%%%%%%%%%%%%%%%%%%%%%%%%%%%%%%%%%%%%%%%%%%%%%%%%%%%%%%%%%%%%%%%%%%%%%%%%%%%%%

Due to its distance \citep[$d$\,= 4-6\,kpc;][]{jones93}, the relatively high wind velocity (40\,${\rm kms}^{-1}$), 
and the remarkable photometric history, \object{IRC\,+10\,420} (= \object{V\,1302~Aql} = 
\object{IRAS\,19244+1115}) is most likely not a post-AGB star evolving through the 
proto-planetary nebula stage as suggested earlier \citep[e.g.][]{fico87,hrivnak89,bokn89,trammell94},
but a yellow hypergiant \citep[$L\sim 5\times 10^5\,L_\odot$, see][]{jones93,oudmaijer96}.
Such extremly luminous stars are extremely rare, and only a dozen are known in the Galaxy \citep{clark05}. 
Yellow hypergiants (YHG) have high mass-loss rates ($10^{-5}-10^{-3} \msunyr$) and are in a short, transitional 
evolutionary stage, thereby rapidly crossing the so-called ``yellow void'' in the Hertzsprung-Russell diagram 
\citep{humphreys02}. Their link to other advanced evolutionary phases of massive stars such as Luminous Blue 
Variables and Wolf-Rayet stars is still an open issue in stellar evolution theory.

Because of its large number of remarkable observational features, \irc has been subject to extensive studies 
over the last 30 years. 
The spectral type of \irc has changed from F8\,Ia$^{+}$  in 1973 \citep{humphreys73} through A5Ia in the mid 90s 
\citep{oudmaijer96,klochkova97} to A2 in 2000 \citep{klochkova02}. Correspondingly, the effective temperature
has changed by more than 3000~K in the last two and a half decades. This makes \irc 
a unique object for the study of stellar evolution since it is one of the very rare stars 
believed to be in the rapid transition from the Red-Supergiant stage to the Wolf-Rayet phase.

HST/WFPC2 images of \ircend's surrounding nebula \citep{humphreys97} revealed a variety of structures including  
condensations or knots, ray-like features, and several arcs or loops within $2^{''}$ from the star, plus one or 
more distant reflection shells. All these features suggest frequent episodes of high mass loss during the past 
centuries. \citet{castro01} found thermal SiO emission in a huge hollow shell around \ircend, with a typical
radius of $\sim10^{17}\,$cm (=6680~AU=1\farcs34 at a distance of 5~kpc), a shell width smaller than half of
the radius, and an expansion velocity of 35\,${\rm kms}^{-1}$.
\citet{castro07} mapped \irc's nebula in the $^{12}{\rm CO}$ J = 2-1 and 1-0 transitions 
and found that the nebula shows an approximately spherical, extended halo surrounding a bright 
inner region,  with both components clearly presenting smaller aspherical features. The CO nebula expands 
isotropically with an expansion rate similar to the SiO shell.

The chemical composition of \ircend's nebula is dominated by O-rich chemistry as suggested by several
observations \citep[e.g.\ the OH maser emission by][]{reid79}, and is similar to O-rich AGB 
stars as recently found by \citet{quintana07}. \irc is also among the brightest IRAS objects and 
one of the warmest stellar OH maser sources known
\citep{giguere76,mutel79,diamond83,bowers84,nebo92}. Ammonia emission has been reported by 
\citet{mcbe80} and \citet{meal95}. From CO observations, high mass-loss rates of the order of 
several $10^{-4}$\,M$_{\odot}{\rm yr}^{-1}$ \citep{knmo85,oudmaijer96,castro07} have been derived.

As inferred by many studies, the structure of the inner circumstellar envelope (CSE; spatial scale of milliarcseconds) 
of \irc appears to be as complex as that of the outer nebula, and numerous scenarios have been proposed to explain its 
observed features. These models include a rotating equatorial disk \citep{jones93}, bipolar outflows \citep{oudmaijer94}, 
infall of circumstellar material \citep{oudmaijer98}, wind blowing in a preferential direction 
\citep{humphreys02}, and even the simultaneous presence of inflowing and outflowing matter \citep{humphreys02}. 

Despite very intensive observing campaigns, the overall geometry of \ircend's CSE is still a matter of debate. 
\citet{humphreys02} obtained HST/STIS spatially resolved spectroscopy of \irc and its reflection nebula. 
They suggested that given the stellar
temperature and the high mass-loss rate, the strong stellar wind of \irc must be optically thick and, thus,
the observed variations in the apparent spectral type and temperature are due to changes in the wind
and do not reflect an interior evolution on short timescales. 
The detection of a nearly spherical CSE by \citet{humphreys02} is in marked contrast to other observations that
reveal a rather axis-symmetric wind geometry in \ircend, such as the earlier HST images \citep{humphreys97},
integral-field spectroscopy \citep{davies07}, or recent spectropolarimetry \citep{patel08}. 
These latter observations suggest a symmetry axis at a position angle of $45^\circ\pm15^\circ$.

Infrared interferometric and coronogra\-phic observations of \irc were reported by, e.g., 
\citet{dyck84}, \citet{ridgway86}, \citet{cofi87}, \citet{christou90},
\citet{kawe95}, \citet{bloecker99}, \citet{sudol99}, \citet{lipman00}, and \citet{mon04}.
\citet{bloecker99} presented
diffraction-limited 73\,mas $K$-band bispectrum speckle-interferometry observations of \ircend's dust shell. 
They found that the $K$-band visibility steeply drops to a plateau-like 
value of $\approx0.6$ at 6~m baseline and, thus, concluded that 40\% of the total $K$-band flux comes
from the extended dust shell. The best radiative transfer model found by \citet{bloecker99} to simultaneously 
explain the spectral energy distribution (SED) and $K$-band visibility 
contains a two-component shell composed of silicate dust with an inner 
rim at 69 stellar radii where the dust temperature is 1000~K.
Moreover, \citet{bloecker99} found that a phase of heavy mass loss with mass-loss rates 
approximately 40 times higher than the current 
$\dot{M}= 1.4\times 10^{-5}\,M_\odot{\rm yr}^{-1}\times(d{\rm kpc}^{-1})$ 
must have ceased roughly 60 to 90 yrs ago.

\citet{mon04} obtained both $K$-band aperture-masking observations of \irc using the Keck I telescope
and long-baseline interferometric observations with the beam-combiner instrument FLUOR at the 
IOTA interferometer \citep[e.g.][]{traub98}. While the aperture masking observations basically confirmed
the results of the speckle observations of \citet{bloecker99}, i.e.\ a sharp visibility drop for baselines
shorter than 2~m and a plateau-like visibility of $\sim$0.6 up to a 8~m baseline, from the IOTA measurements 
\citet{mon04} found a $K$-band visibility of $\sim$0.7, from which they concluded that the compact stellar
component is not resolved at baselines as long as $\sim$35~m.

Recently, \citet{dewit08} presented the first near-infrared long-baseline interferometric observations of
\irc obtained with the AMBER instrument at ESO's Very Large Telescope Interferometer (VLTI).
The observations presented by \citet{dewit08} were carried out in medium spectral resolution mode 
($\lambda/\Delta\lambda=1\,500$) around the \brg emission line. \citet{dewit08} resolved the \brg line-emitting 
region and derived a Gaussian FWHM size of 3.3~milliarcseconds (mas), 
but due to calibration problems, the size of the continuum-emitting region could not be constrained. 

In this paper, we present the first VLTI/AMBER observations of \irc in the $H$ and
$K$ bands in low-spectral resolution mode and the first AMBER measurements of its \brge-emitting region
with a spectral resolution of $\lambda/\Delta\lambda=12\,000$ (high-spectral resolution mode of AMBER)
and baselines up to 128~m. From these new AMBER observations, the size of the continuum-emitting region 
in several spectral channels across the $H$ and $K$ bands could be derived and, by comparison of the 
high-spectral resolution AMBER data with 2-D gas radiative transfer models, details on the size and 
geometry of the \brg line-emitting region could be obtained.

The observations presented here are the first step of an observing campaign to better constrain the
geometry of the inner wind region of \irc at a scale of a few stellar radii.

The paper is structured as follows. In Sect.~\ref{sect_obs} we will present the AMBER observations of \irc
and discuss the data reduction. In Sect.~\ref{sect_res_fov} we discuss the implications
of AMBER's limited field-of-view for the observations of the very extended object \ircend.
In Sect.~\ref{sect_res_lr} and \ref{sect_res_hr}, the absolute calibration, radiative transfer modeling, 
and interpretation of the AMBER data will be discussed. 
The paper closes with a summary and conclusions in Sect.~\ref{sect_concl}.

%%%%%%%%%%%%%%%%%%%%%%%%%%%%%%%%%%%%%%%%%%%%%%%%%%%%%%%%%%%%%%%%%%%%%%%%%%%%%%%%%%%%%%%%%%%%%%%%%%%%%%%%%%
\section{Observations and data reduction}\label{sect_obs}
%%%%%%%%%%%%%%%%%%%%%%%%%%%%%%%%%%%%%%%%%%%%%%%%%%%%%%%%%%%%%%%%%%%%%%%%%%%%%%%%%%%%%%%%%%%%%%%%%%%%%%%%%%

%%%%%%%%%%%%%%%%%%%%%%%%%%%%%%%%%%%%%%%%%%%%%%%%%%%%%%%%%%%%%%%%%%%%%%%%%%%%%%%%%%%%%%%%%%%%%%%%%%%%%%%%%%
\subsection{General remarks on the data reduction process}\label{sect_amdlib}
%%%%%%%%%%%%%%%%%%%%%%%%%%%%%%%%%%%%%%%%%%%%%%%%%%%%%%%%%%%%%%%%%%%%%%%%%%%%%%%%%%%%%%%%%%%%%%%%%%%%%%%%%%

In the following two subsections we describe our VLTI/AMBER observations of \irc in low (LR) and high spectral 
resolution (HR) mode. AMBER is the near-infrared closure-phase beam combiner of ESO's VLTI, simultaneously
operating in the $J$, $H$, and $K$ bands. This instrument is capable of interfering the light of either 
three 1.8~m Auxiliary Telescopes (ATs) or three of the 8.2~m Unit Telescopes (UTs) located on Paranal. 
A detailed description of the AMBER instrument and its optical design is given in \citet{petrov07} and 
\citet{robbe07}. The AMBER data presented here were reduced using version 2.2 of the data reduction package 
{\it amdlib}
\footnote{This software package is kindly provided by the Jean-Marie Mariotti Center and publicly available 
from http://www.jmmc.fr/data\_processing\_amber.htm.}. The reduction software is based on the so-called
Pixel-to-Visibility-Matrix (P2VM) algorithm, which is described in \citet{tatulli07}.

Following previous experiences with the reduction
of low and high-spectral resolution AMBER data \citep[e.g.][]{weigelt07,wittkowski08}, the data selection was 
carried as follows: As primary selection criteria we used the signal-to-noise ratio (SNR) of the fringe signal.
For the LR data, we kept 20\% of the data with the highest fringe SNR, while for the HR data taken with the
fringe tracker FINITO \citep[e.g.\ ][]{gai03,jbb08}
we kept 80\% of the data with the highest fringe SNR. In both cases, the optimal fringe SNR selection value 
was found by increasing the fraction of discarded data frames from 0 to 80\%. 
A stronger selection than given above only led to an increase 
of the noise; i.e., finally there was only a decrease of the data quality rather than a significant improvement. 
In addition to the fringe SNR selection, to ensure a proper deselection of all frames which are far 
from zero optical path difference, we also discarded in the low- as well as 
high-spectral resolution data all frames with a piston of more than $10\,\mu$m. For the high-spectral resolution
data, this seems to be a too strong selection, since in this case the coherence length is of the order of 
$\lambda/\Delta\lambda\times\lambda\approx2$~cm, 
but due to the use of FINITO the piston variation is typically much small than 100 micron in the HR data. 
Since it turned out that
the HR results do not show a strong variation with piston selection, we finally decided to use the same piston 
selection criteria for both the HR and LR data.

Concerning the error estimates of our AMBER measurements, we were facing the problem that for each observation
only one suitable calibrator measurement was available. Thus, in addition to the statistical error of the
observables calculated from the average over all frames of a given data set, it was problematic to address also any 
systematic error sources. Therefore, to account for systematic uncertainties of the absolute calibration process, 
we looked up previous AMBER measurements taken under similar weather conditions. From these comparisons, we concluded 
to add an absolute error of 0.03 to all visibility data and 3 degrees to all phases to account for the systematic
uncertainties.

%%%%%%%%%%%%%%%%%%%%%%%%%%%%%%%%%%%%%%%%%%%%%%%%%%%%%%%%%%%%%%%%%%%%%%%%%%%%%%%%%%%%%%%%%%%%%%%%%%
%%%
%%% Fig.1: uv coverage
%%%
%%%%%%%%%%%%%%%%%%%%%%%%%%%%%%%%%%%%%%%%%%%%%%%%%%%%%%%%%%%%%%%%%%%%%%%%%%%%%%%%%%%%%%%%%%%%%%%%%%
   \begin{figure}
	\vspace*{-1mm}

   \centering
   \includegraphics[angle=-90,width=0.3\textwidth]{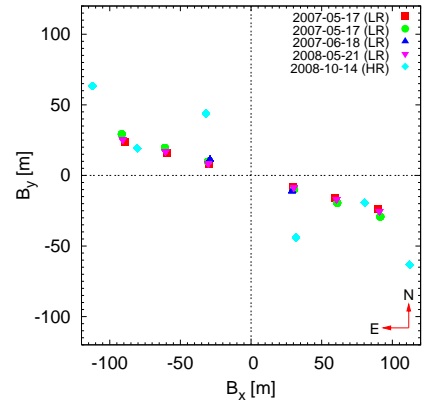}
      \caption{($u,v$) coverage of the VLTI/AMBER observations of \irc in low (LR) and high (HR) 
                       spectral resolution mode presented in this paper.
              }
         \label{fig1_uv}
	\vspace*{-3mm}

   \end{figure}
%%%%%%%%%%%%%%%%%%%%%%%%%%%%%%%%%%%%%%%%%%%%%%%%%%%%%%%%%%%%%%%%%%%%%%%%%%%%%%%%%%%%%%%%%%%%%%%%%%%

%%%%%%%%%%%%%%%%%%%%%%%%%%%%%%%%%%%%%%%%%%%%%%%%%%%%%%%%%%%%%%%%%%%%%%%%%%%%%%%%%%%%%%%%%%%%%%%%%%%%%%%%%
\subsection{The low-spectral resolution data}\label{sect_obs_lr}
%%%%%%%%%%%%%%%%%%%%%%%%%%%%%%%%%%%%%%%%%%%%%%%%%%%%%%%%%%%%%%%%%%%%%%%%%%%%%%%%%%%%%%%%%%%%%%%%%%%%%%%%%%

The low-spectral resolution (hereafter LR) AMBER observations of \irc were obtained with the Auxiliary 
Telescopes (ATs) in May/June 2007 and May 2008 using the linear baseline configurations E0-G0-H0 and A0-D0-H0.
A summary of the LR observations is given in Table~\ref{tab1_lrdata}.
As this table as well as the $(u,v)$ coverage in Fig.~\ref{fig1_uv} shows, the total baseline and position angle 
ranges covered by the LR observations of \irc 
are 15 to 94~m and 69 to 75\degr, respectively, i.e.\ the observations
were essentially carried out in the same direction on the sky. All measurements were made under sub-average to
average seeing conditions (see Tabs.~\ref{tab1_lrdata} and ~\ref{tab3_hrdata}). For the calibration of all LR 
AMBER data on \irc we used the calibrator HD~190327. A record of the calibrator observations is also provided
in Table~\ref{tab1_lrdata}.

%%%%%%%%%%%%%%%%%%%%%%%%%%%%%%%%%%%%%%%%%%%%%%%%%%%%%%%%%%%%%%%%%%%%%%%%%%%%%%%%%%%%%%%%%%%%%%%%%%%%%%%%%%
%%%
%%% Tab.1: AMBER LR observations
%%%
%%%%%%%%%%%%%%%%%%%%%%%%%%%%%%%%%%%%%%%%%%%%%%%%%%%%%%%%%%%%%%%%%%%%%%%%%%%%%%%%%%%%%%%%%%%%%%%%%%%%%%%%%%
   \begin{table}[h]
      \caption[]{
       Summary of the AMBER low-spectral resolution observations of \object{IRC+10420} 
       and the calibrator star HD~190327. UTC is the time stamp (Universal Time)
       corresponding to the start of fringe tracking, $B_{\rm p}$ is the projected baseline, and PA denotes 
       the position angle of the observation. The seeing value given in the last column corresponds to the
       DIMM seeing at the time of the corresponding observation. Each measurement consists of 5000 frames 
       with an exposure time of 50~ms. The calibrator HD~190327 has spectral type K0III, $K=3.2^{\rm m}$
       \citep{2mass}, and a uniform-disk diameter $d=1.10\pm0.01$~mas \citep{charm2}.
       }
         \label{tab1_lrdata}
	\vspace*{-5mm}
     $$ 
         \begin{array}{cccccc}
            \hline
            \noalign{\smallskip}
            {\rm date}& {\rm UTC}& {\rm AT}           & B_{\rm p}        & PA      & {\rm seeing}\\
                      &          & {\rm configuration}& [{\rm m}]        & [\degr] & [\arcsec]   \\
             \noalign{\smallskip}
            \hline
            \noalign{\smallskip}
	\multicolumn{6}{c}{{\rm \irc}}\\ \hline
	 2007-05-17   & 08:11:53  & {\rm A0-D0-H0}    &31-93             & 75      & 1.0\\
	 2007-05-17   & 09:18:51  & {\rm A0-D0-H0}    &32-96             & 72      & 1.4\\
	 2007-06-18   & 08:09:22  & {\rm E0-G0-H0}    &15-48^{\mathrm{a}}& 69      & 0.7\\
	 2008-05-21   & 08:12:14  & {\rm A0-D0-H0}    &31-94             & 74      & 0.8\\ \hline
	\multicolumn{6}{c}{{\rm HD~190327}}\\ \hline
         2007-05-17   & 08:48:23  & {\rm A0-D0-H0}    &31-93             & 74      & 1.2 \\
	 2007-06-18   & 08:39:13  & {\rm E0-G0-H0}    &16-47             & 70      & 0.6 \\
	 2008-05-21   & 08:48:41  & {\rm A0-D0-H0}    &32-95             & 74      & 0.7 \\
	
            \noalign{\smallskip}
            \hline
         \end{array}
     $$ 
\begin{list}{}{}
\item[$^{\mathrm{a}}$] Due to technical problems during the observations, visibilities could only 
                       be obtained for the intermediate baseline ($B_{p}=31$~m).
\end{list}
	\vspace*{-5mm}

   \end{table}
%%%%%%%%%%%%%%%%%%%%%%%%%%%%%%%%%%%%%%%%%%%%%%%%%%%%%%%%%%%%%%%%%%%%%%%%%%%%%%%%%%%%%%%%%%%%%%%%%%%%%%%%%%

%%%%%%%%%%%%%%%%%%%%%%%%%%%%%%%%%%%%%%%%%%%%%%%%%%%%%%%%%%%%%%%%%%%%%%%%%%%%%%%%%%%%%%%%%%%%%%%%%%%%%%%%%%
%%%
%%% Fig.2: AMBER LR observables
%%%
%%%%%%%%%%%%%%%%%%%%%%%%%%%%%%%%%%%%%%%%%%%%%%%%%%%%%%%%%%%%%%%%%%%%%%%%%%%%%%%%%%%%%%%%%%%%%%%%%%%%%%%%%%
   \begin{figure*}
   \centering
   \includegraphics[angle=0,width=1.0\textwidth]{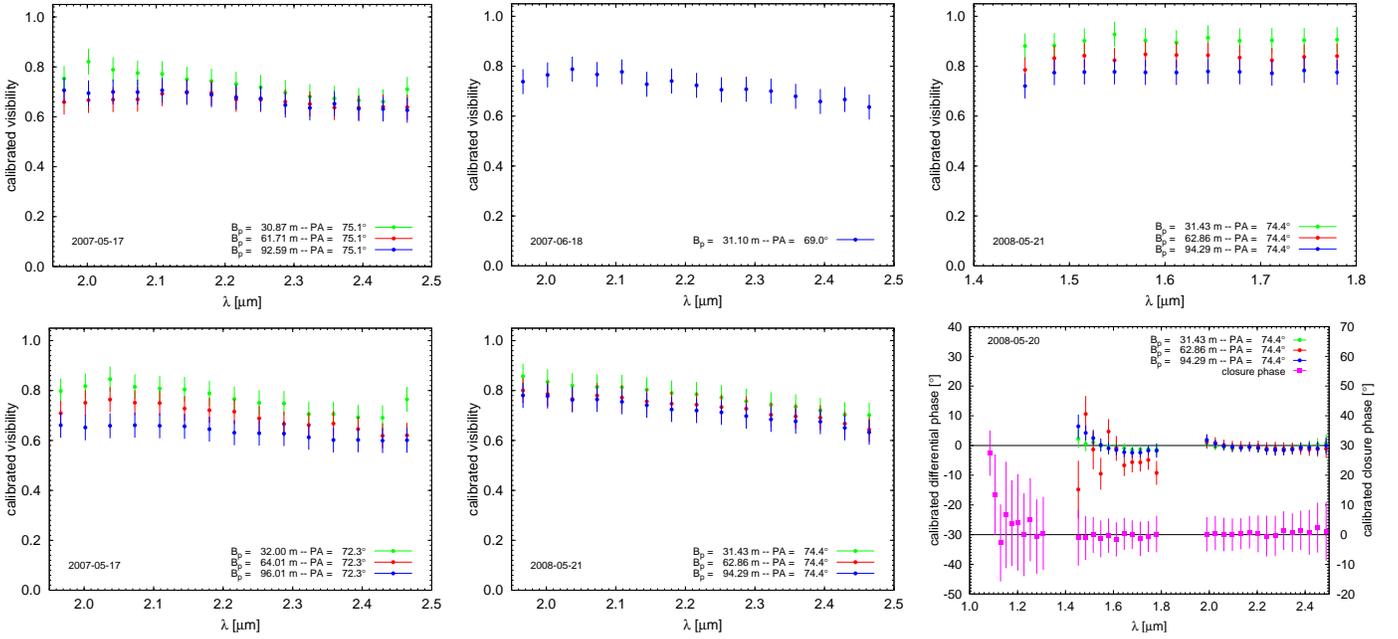}
      \caption{VLTI/AMBER observations of \irc in low-spectral resolution mode. 
       The panels in the left and middle columns show the calibrated $K$-band visibility
       as a function of wavelength for the observations listed in Table~\ref{tab1_lrdata}. 
       The top right panel shows the calibrated $H$-band visibility of \irc as a function of wavelength 
       from the May 2008 observations. As will be outlined in \ref{sect_res_fov} in more detail, all visibilities
       suffer from a strong field-of-view effect.
% since AMBER's limited FOV leads to a significant truncation
%       of \ircend's extended dust shell, resulting in considerably higher visibilities.
       Finally, the bottom right panel presents the differential phase
       in the $H$ and $K$ bands (left y-axis scale) and the closure phase (right y-axis scale) in the 
       $J$, $H$, and $K$ bands as a function of wavelength for the observations carried out in May 2008. 
%       As the panel reveals, for both the differential and closure phases no significant non-zero phase signal 
%       could be detected within the uncertainty limits.
              }
         \label{fig2_lrdata}
   \end{figure*}
%%%%%%%%%%%%%%%%%%%%%%%%%%%%%%%%%%%%%%%%%%%%%%%%%%%%%%%%%%%%%%%%%%%%%%%%%%%%%%%%%%%%%%%%%%%%%%%%%%%%%%%%%%

Unfortunately, due to technical problems during the observations, from the measurements carried
out in June 2007 reliable data could only be retrieved from one of the three baselines. Thus, in this case
only one visibility instead of a triplet was obtained. Due to unfavorable weather conditions and limitations
of the technical performance, reliable differential and closure phases from the LR observations could not
be derived for any of the 2007 observations, but only for those from May 2008.

The results of the AMBER LR observations of \irc are shown in Fig.~\ref{fig2_lrdata}. The panels in the left and
middle column show the AMBER $K$-band visibilities as a function of wavelength for the three epochs covered by our 
study, the top right panel shows the the AMBER $H$-band visibilities as a function of wavelength, and the bottom right
panel illustrates the differential and closure phases in the $J,H$, and $K$ bands obtained from the May 2008
observations of \ircend. 

Two main results can be seen in Fig.~\ref{fig2_lrdata}. First, both the $H$- and
$K$-band visibilities show only a weak wavelength dependence, and the visibility change with baseline for a
given wavelength is only moderate. This corresponds to the fact that even with the shortest baselines of
our AMBER observations, the extended dust shell of \irc is already fully resolved, and thus, AMBER probes
the compact stellar component, i.e.\ the continuum emission from the central star and the dense circumstellar
wind. The second result from the calibrated LR observables shown in Fig.~\ref{fig2_lrdata} is that both the
differential and closure phases show zero phase signals across all three near-infrared bands
within the uncertaintly limits. Thus, in the LR AMBER data which mainly probe the continuum emission 
\irc does not show detectable deviations from point symmetry.

%%%%%%%%%%%%%%%%%%%%%%%%%%%%%%%%%%%%%%%%%%%%%%%%%%%%%%%%%%%%%%%%%%%%%%%%%%%%%%%%%%%%%%%%%%%%%%%%%%%%%%%%%%
\subsection{The high-spectral resolution data}\label{sect_obs_hr}
%%%%%%%%%%%%%%%%%%%%%%%%%%%%%%%%%%%%%%%%%%%%%%%%%%%%%%%%%%%%%%%%%%%%%%%%%%%%%%%%%%%%%%%%%%%%%%%%%%%%%%%%%%

In addition to the low-spectral resolution $H$- and $K$-band AMBER observations of \irc presented in the
previous section, we also obtained the first AMBER measurement of \irc in high-spectral resolution mode
(spectral resolution $\lambda/\Delta\lambda = 12\,000$). The observations
were carried out in October 2008 under average seeing conditions as part of an AMBER science verification 
run in order to test the performance of
AMBER using the UTs and the fringe tracker FINITO. In the context of this science verification 
run, a single measurement of \irc and the calibrator star HD~232078 was carried out in the $K$ band, centered
around the \brg emission line ($\lambda_{\rm c}= 2.166\,\mu$m), with an exposure time of 3~s. Due to this
long exposure time, a window covering 512 pixels, i.e.\ all spectral channels of the AMBER detector, 
could be read out, corresponding to a wavelength coverage between $\sim 2.145$ and $\sim 2.19\,\mu$m.

%%%%%%%%%%%%%%%%%%%%%%%%%%%%%%%%%%%%%%%%%%%%%%%%%%%%%%%%%%%%%%%%%%%%%%%%%%%%%%%%%%%%%%%%%%%%%%%%%%%%%%%%%%
%%%
%%% Tab.2: AMBER HR observations
%%%
%%%%%%%%%%%%%%%%%%%%%%%%%%%%%%%%%%%%%%%%%%%%%%%%%%%%%%%%%%%%%%%%%%%%%%%%%%%%%%%%%%%%%%%%%%%%%%%%%%%%%%%%%%
%%%%%%%%%%%%%%%%%%%%%%%%%%%%%%%%%%%%%%%%%%%%%%%%%%%%%%%%%%%%%%%%%%%%%%%%%%%%%%%%%%%%%%%%%%%%%%%%%%%%%%%%%%
   \begin{table}
      \caption[]{
       Summary of the AMBER observations of \object{IRC+10420} and the calibrator star HD~232078
       in high-spectral resolution mode from Oct 14, 2008 using the UT telescope configuration UT1-UT2-UT4. 
       All measurements were carried out using the fringe tracker FINITO and with an exposure time of 3~s. 
       The uniform-disk diameter of HD~232078 (spectral type K3IIp), $d_{\rm UD} = 0.74\pm0.1$~mas,
       was taken from the CHARM2 catalogue \citep{charm2}. $N_{\rm frames}$ is the total number of recorded frames.
       For the meaning of the other columns, see Table~\ref{tab1_lrdata}.
       }
         \label{tab3_hrdata}
	\vspace*{-5mm}
     $$ 
         \begin{array}{lccccc}
            \hline
            \noalign{\smallskip}
            {\rm object}& {\rm UTC}& N_{\rm frames}& B_{\rm p}   & PA      & {\rm seeing}\\
                      &            &               & [{\rm m}]   & [\degr] & [\arcsec]   \\
             \noalign{\smallskip}
            \hline
            \noalign{\smallskip}
	    {\rm \irc}      & 00:26:40  & 125            & 54-129      & 36-75   & 0.9\\
	    {\rm HD~232078} & 00:03:59  & 125            & 51-130      & 37-78   & 1.1\\
            \noalign{\smallskip}
            \hline
         \end{array}
     $$ 
	\vspace*{-5mm}

   \end{table}
%%%%%%%%%%%%%%%%%%%%%%%%%%%%%%%%%%%%%%%%%%%%%%%%%%%%%%%%%%%%%%%%%%%%%%%%%%%%%%%%%%%%%%%%%%%%%%%%%%%%%%%

It should be noted that a second measurement of \irc with an exposure time of 1~s was discarded for
two reasons. First of all, the data quality of this second measurement is lower than that with the longer
detector integration time (DIT), simply because the SNR is lower. Second, there is no corresponding calibrator
measurement with the same DIT close in time to the science target measurement. From the ESO data archive,
we found two other calibrator measurements with DIT=$1\,$s from different science programs in the same night,
but these measurements of the calibrators HD~902 and HD~13692 were taken 1 and 2 hours after the 
observations of \ircend. Since current high-spectral resolution AMBER data suffer from a time-variable
high-frequency beating introduced by the VLTI Infrared Image Sensor IRIS \citep{iris}, we decided to discard these 
additional measurements because a proper calibration of the science target data turned out to be impossible.
This seems to be a general problem for observations in high-spectral resolution mode if the time 
span between the science target and calibrator measurement becomes longer than approximately 30-45~minutes.
For the same reason, we also discarded two other calibrator data sets with the same exposure time (DIT=3~s)
as the measurement of \irc presented here since these calibrator observations were carried out 
$\sim5.5$ and 8 hours after the \irc observations. To further reduce the effect of the IRIS fringes in our
data, we spectrally binned our final visibilities and phases with a bin size of 7 and a step size of 1. To avoid
a loss of spectral resolution across the \brg line, this binning was only performed in the continuum region, i.e.\ 
for wavelength channels with $\lambda < 2.1660\,\mu$m or $\lambda > 2.1677\,\mu$m.
A summary of the HR observations discussed in this paper is given in Table~\ref{tab3_hrdata}.

%%%%%%%%%%%%%%%%%%%%%%%%%%%%%%%%%%%%%%%%%%%%%%%%%%%%%%%%%%%%%%%%%%%%%%%%%%%%%%%%%%%%%%%%%%%%%%%%%%%%%%%%%%
%%%
%%% Fig.3: AMBER HR observables
%%%
%%%%%%%%%%%%%%%%%%%%%%%%%%%%%%%%%%%%%%%%%%%%%%%%%%%%%%%%%%%%%%%%%%%%%%%%%%%%%%%%%%%%%%%%%%%%%%%%%%%%%%%%%%
   \begin{figure*}
	\vspace*{-5mm}

   \centering
   \includegraphics[angle=-90,width=0.75\textwidth]{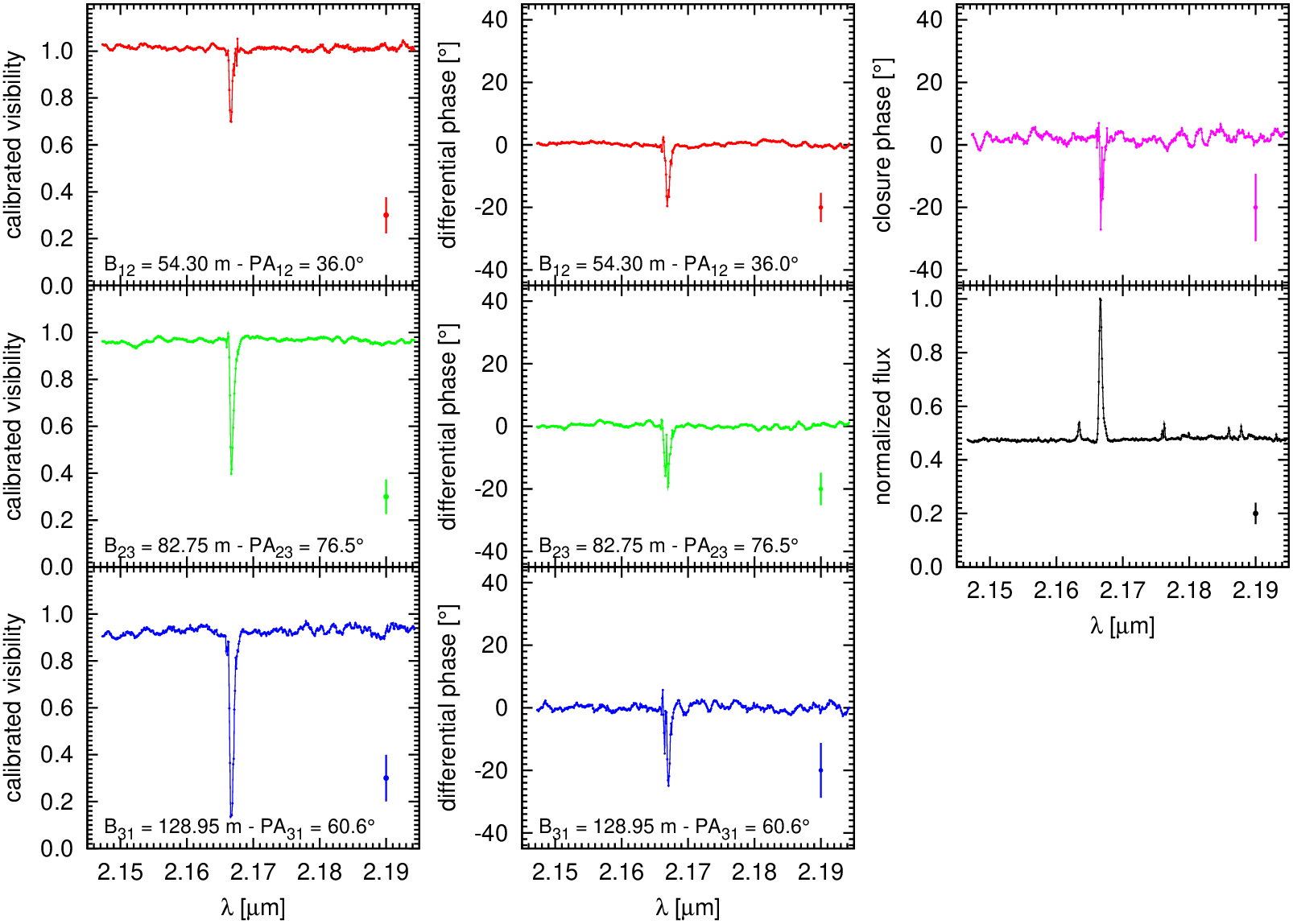} \\
   \includegraphics[angle=-90,width=0.75\textwidth]{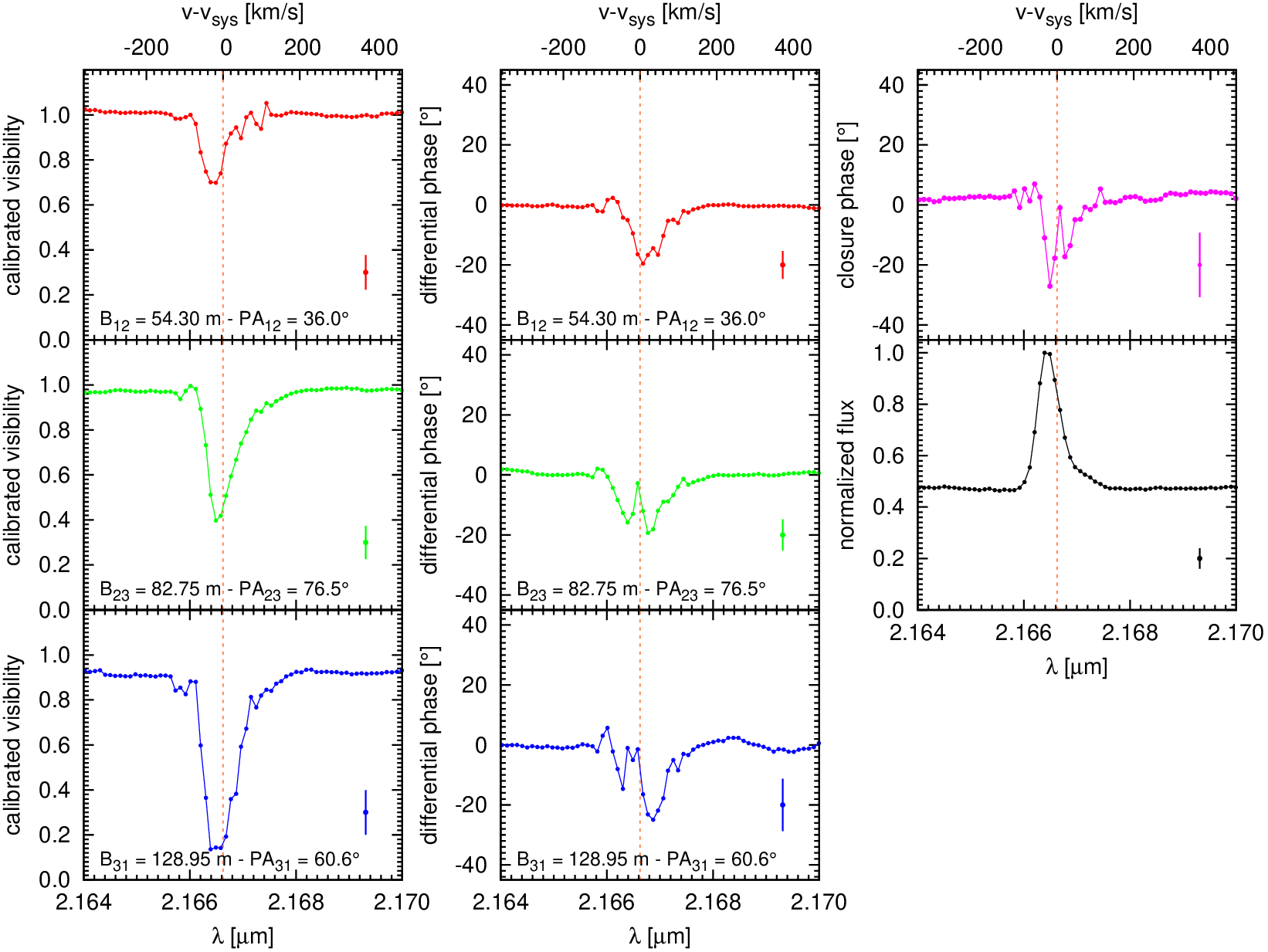}
      \caption{VLTI/AMBER observations of \irc in high-spectral resolution mode around the \brg emission line. 
       The two left panels show the calibrated visibility and differential phase
       as a function of wavelength for all three two-telescope baselines, while the panels on the right
       show the calibrated closure phase (top) and the normalized spectrum, i.e.\ the spectrum of \irc divided by
       the calibrator spectrum and normalized to peak intensity. 
       While the top panels show the observables across the full wavelength range covered 
       by the AMBER measurements, all bottom panels show the same quantities in a wavelength region close to the 
       \brg emission-line in more detail. The visibilities and phases in the
       continuum region ($\lambda < 2.1660\,\mu$m and $\lambda > 2.1677\,\mu$m) have been spectrally binned 
       to reduce the high-frequency beating introduced by IRIS. The absolute wavelength calibration was performed
       by comparing the raw spectra of all data sets with telluric spectra from the Kitt Peak Observatory,
       which were convolved to match the spectral resolution of the AMBER measurements.
              }
         \label{fig3_hrdata}
   \end{figure*}
%%%%%%%%%%%%%%%%%%%%%%%%%%%%%%%%%%%%%%%%%%%%%%%%%%%%%%%%%%%%%%%%%%%%%%%%%%%%%%%%%%%%%%%%%%%%%%%%%%%%%%%%%%

The calibrated observables of the AMBER high-spectral resolution measurements are presented in Fig.~\ref{fig3_hrdata}.
The panels in the three left columns show (from left to right) the visibilities, differential phases, closure phases 
(top), and the spectrum (bottom) for the full wavelength range covered by the observations. 
All panels in the three columns on the right-hand side of Fig.~\ref{fig3_hrdata} show the same quantities as the 
corresponding panels in the three left columns, but only for a small wavelength range close to the \brg emission line.
For the sake of clarity, in each panel, only a single averaged error bar is shown on the right-hand side. 

The wavelength
calibration of the HR data was performed by comparison of the AMBER raw spectra of both \irc and the calibrator
HD~232078 with high-spectral resolution ($\lambda/\Delta\lambda=60\,000$) telluric spectra kindly provided by
the Kitt Peak Observatory\footnote{The spectra are retrievable from\\
http://www.eso.org/sci/facilities/paranal/instruments/isaac/tools/spec\-tros\-copic\_standards.html\#Telluric.}.
The vertical dashed line shown in all panels on the right-hand side of Fig.~\ref{fig3_hrdata} marks the zero velocity 
with respect to earth, assuming a radial system velocity of 73\,${\rm kms}^{-1}$ as given by 
\citet{humphreys02}, and a heliocentric correction of -26\,${\rm kms}^{-1}$ for the time of the observations. 
The spectrum shown in Fig.~\ref{fig3_hrdata} is the average over the photometric beams
from the three single telescopes and the interferometric signal.

The calibration of the high-spectral resolution AMBER measurements is more
uncertain compared to the low-spectral resolution data shown in Sect.~\ref{sect_obs_lr} for two reasons. First of
all, for the HR data, we had only one calibrator measurement instead of three as in the case of the LR data. 
Second, the HR data were recorded with the fringe tracker FINITO. Since the performance of FINITO depends on, e.g.,
a target's correlated flux and the weather conditions at the time of the observations, the final calibrated 
observables obtained
from AMBER+FINITO observations can be biased either due to a significant brightness difference between science
target and calibrator and/or due to different weather conditions (seeing, coherence time, wind) during the 
science target and calibrator measurements.

In the case of the HR measurements of \irc presented in this paper, the calibration relies on only a single
calibrator measurement of HD~23078 which is approximately 0.5 magnitudes fainter in the $K$ band than \irc
($K_{\rm HD23078} = 4.2^{\rm m}; K_{\rm IRC+10420}=3.65^{\rm m}$). From the weather 
data\footnote{Apart from the information provided by ESO with the raw data itself, the ambient weather conditions 
during an observation can be looked up at http://archive.eso.org/asm/ambient-server.}, 
we can see that at the time of the calibrator and science target measurements the seeing was varying between
0\farcs85 and 1\farcs15, and the wind speed on the ground was steadily declining from 9 to 7.5\,${\rm ms}^{-1}$. 
Thus, there is a noticeable, but not dramatic change in the weather conditions between the calibrator and \irc 
observation. 
Together with the 0.5 magnitude brightness difference, this might explain why, for instance, the calibrated 
visibility at the shortest baseline (see Fig.~\ref{fig3_hrdata}) is slightly larger than unity. 
In fact, when using data from other calibrators of the same night (HD~25680; HD~33833) which were taken 
several hours later and with a seeing of 1\farcs4, the calibrated visibility of \irc reaches values $> 1.2$.

While the weaker emission peaks seen in the full AMBER spectrum of \irc in Fig.~\ref{fig3_hrdata} are telluric 
artifacts of the calibration process and therefore do not indicate real emission features, the spectrum clearly
shows a strong \brg emission line arising from \irc which contributes approximately 50\% to the total flux 
at $\lambda=2.1666\,\mu$m. The \brg line is clearly blueshifted with $v=-25\,{\rm kms}^{-1}$ 
with respect to the systemic velocity, in agreement with previous findings by, e.g., \citet{oudmaijer94} 
($v=-30\,{\rm kms}^{-1}$) and \citet{humphreys02} ($v=-22\,{\rm kms}^{-1}$).
From the $\lambda$-calibration process of our data, we estimate that the uncertainty of the
wavelength calibration is of the order of $10\,{\rm kms}^{-1}$.
The equivalent width of the \brg line in the AMBER spectrum is $-6.7~\AA$, in good agreement 
with the value found by \citet{dewit08} from medium-spectral resolution AMBER observations in June 2006.
The \brg line is slightly asymmetric in the sense that the redshifted tail of the line is more pronounced
with velocities up to $\sim120\,{\rm kms}^{-1}$ compared to $\sim80\,{\rm kms}^{-1}$ on the blueshifted side.

At the wavelengths of the \brg emission, the visibilities on all three baselines show a strong decrease, indicating
that the line-emitting zone is clearly resolved by our AMBER observations and more extended than the 
continuum-emitting region. At the longest baseline, the visibility in the center of the \brg line is as low as 
0.14. It should be noted here that similar to the LR data presented in Sect.~\ref{sect_obs_lr} the HR data also 
suffer from a strong field-of-view (FOV) effect as will be discussed in more detail in the next section. 

The differential phases in the region of the \brg-emission line show a clear non-zero signal within the 
uncertainty limits. On all three baselines, we obtained the strongest phase signal in the redshifted wing
of the \brg-emission line, with phases up to $\phi_{ij}=-20^\circ$ on the two shorter baselines and 
$\phi_{13}=-26^\circ$ on the longest baseline. 
The closure phase $\Phi$, on the other hand, is in good agreement with the differential closure phase
$\Phi_{\rm diff} = \phi_{12} + \phi_{23} - \phi_{13}$. $\Phi$ also shows a strong signal in the redshifted
wing of the \brg-emission line ($\sim-18^\circ$) and the strongest non-zero signal ($\sim-27^\circ$) at the 
line center.

%%%%%%%%%%%%%%%%%%%%%%%%%%%%%%%%%%%%%%%%%%%%%%%%%%%%%%%%%%%%%%%%%%%%%%%%%
\section{Results and discussion}\label{sect_res}
%%%%%%%%%%%%%%%%%%%%%%%%%%%%%%%%%%%%%%%%%%%%%%%%%%%%%%%%%%%%%%%%%%%%%%%%%
%%%%%%%%%%%%%%%%%%%%%%%%%%%%%%%%%%%%%%%%%%%%%%%%%%%%%%%%%%%%%%%%%%%%%%%%%
\subsection{The field-of-view effect}\label{sect_res_fov}
%%%%%%%%%%%%%%%%%%%%%%%%%%%%%%%%%%%%%%%%%%%%%%%%%%%%%%%%%%%%%%%%%%%%%%%%%

Since VLTI/AMBER is a single-mode fiber instrument, the field-of-view (FOV) is limited to the Airy disk 
of the telescope aperture. Therefore, for AMBER observations with the 1.8~m ATs of the VLTI,
the FOV is 250~mas in the $K$ band, while for AMBER observations with the 8.2~m UTs, the FOV is 
only $\sim60$~mas. Thus, when observing an object as extended as \irc with AMBER, it is of great importance
to take AMBER's limited FOV into account, since a non-negligible fraction of the total flux will be located
well outside the FOV of the observations.

%%%%%%%%%%%%%%%%%%%%%%%%%%%%%%%%%%%%%%%%%%%%%%%%%%%%%%%%%%%%%%%%%%%%%%%%%%%%%%%%%%%%%%%%%%%
%%%
%%% Fig.4: comparison of speckle and AMBER results/FOV-effect
%%%
%%%%%%%%%%%%%%%%%%%%%%%%%%%%%%%%%%%%%%%%%%%%%%%%%%%%%%%%%%%%%%%%%%%%%%%%%%%%%%%%%%%%%%%%%%%
   \begin{figure*}[ht]
   \vspace*{-0mm}

   \includegraphics[angle=0,width=\textwidth]{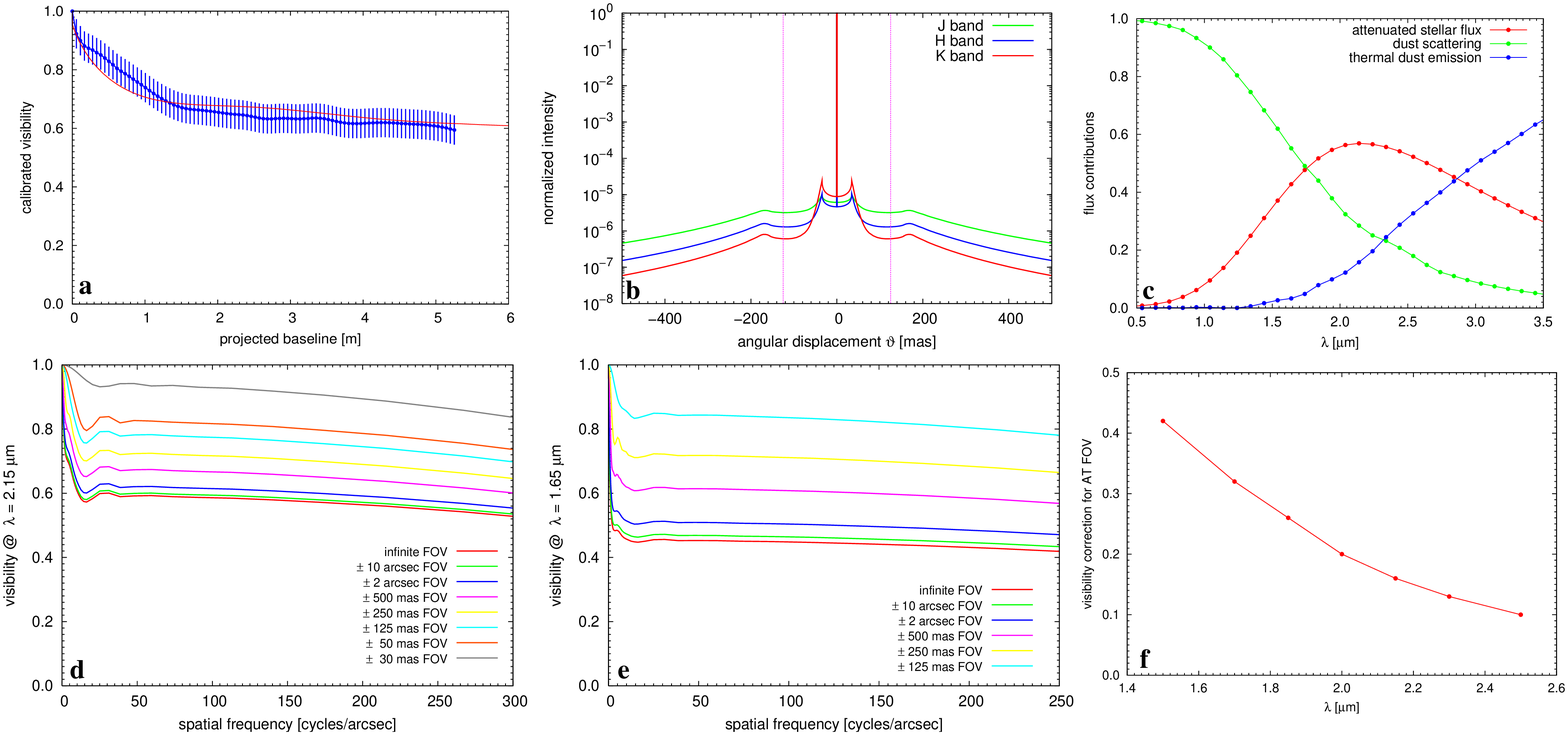}
      \caption{
	{\it Top row}:
	{\it a}:
        Azimuthally averaged $K$-band visibility ($\lambda=2.11\,\mu$m) of \irc obtained
       	from speckle-interferometric observations with the SAO 6~m telescope \citep[][blue bullets with errorbars]{
        bloecker99}. Up to the cutoff-frequency, the visibility drops to a plateau value of $\sim$0.6. 
        The red solid line shows the best-fitting radiative transfer model of \citet{bloecker99} 
        obtained with the 1-D code DUSTY. This model is able to reproduce both 
        the spectral energy distribution and the $K$-band visibility.
	{\it b}: 
        Near-infrared intensity profiles of the best-fitting DUSTY model 
        of \citet{bloecker99} The profiles are shown for $\lambda=1.25, 1.65,$ and $2.2\,\mu$m, corresponding
        to the central wavelengths in the $J$, $H$, and $K$ bands. All curves show a characterstic limb-brightening
        at the inner edge of the dust shell. The dashed vertical lines indicate AMBER's field-of-view (FOV)
        in the case of observations with the 1.8~m-AT telescopes ($\pm 125$~mas). Note that for the high-spectral
        resolution AMBER observations with the UT telescopes, the FOV is only $\pm 30$~mas.
	{\it c}:
        Relative contributions of attenuated stellar flux (red), scattered light (green), and thermal dust emission 
        (blue) as a function of wavelength as inferred by the final DUSTY model of \citet{bloecker99} for an
        infinite FOV.
	{\it Bottom row}:
	{\it d,e}:
	Visibility from the best-fitting DUSTY model of \citet{bloecker99} as a function of spatial frequency for 
        $\lambda=2.11$ (panel d) and $1.65\,\mu$m (panel e). The different curves show the visibility obtained with 
        different FOVs (see labels). 
	{\it f}:
        AMBER visibility correction for \irc as a function of wavelength across the $H$ and $K$ bands according to 
        the DUSTY model of \citet{bloecker99}. The correction is given for a 250~mas FOV, which corresponds to AMBER
        observations with the ATs. See text for further discussions.
        }

         \label{fig4_foveffect}
   \vspace*{-0mm}

   \end{figure*}
%%%%%%%%%%%%%%%%%%%%%%%%%%%%%%%%%%%%%%%%%%%%%%%%%%%%%%%%%%%%%%%%%%%%%%%%%%%%%%%%%%%%%%%%%%%%%%%%%%%%%%%%%%%%%%%%%%%%%

To account for the FOV effect in the AMBER data of \ircend, we used the dust radiative 
transfer model of \citet{bloecker99}. This radiative transfer model was developed with the code
DUSTY \citep{dusty,dusty99,dusty00} to simultaneously explain the spectral energy distribution (SED)
and $K$-band speckle-interferometric observations obtained with the SAO 6~m telescope. 

The azimuthally averaged $K$-band
visibility of \citet{bloecker99} is shown in the top-left panel of Fig.~\ref{fig4_foveffect} as a function of
baseline length. As one can see, the steep visibility drop ends in a plateau at a visibility level of $\sim$60\%.
The plateau itself is associated with the unresolved stellar component plus the dense stellar wind of \ircend, 
which can be probed by interferometry with baselines longward of approximately 10~m. On the other hand, the 
extended dust shell is fully resolved already with a few meter baseline.
According to the best-fitting model of Bl\"ocker et al., as a result of its heavy mass loss 
\irc is surrounded by an optically thick dust shell 
composed of silicate dust, which contributes approximately 40\% to the total $K$-band flux of \ircend.
The inner boundary of the dust shell which marks the dust sublimation radius 
is located at $69~R_{\star}\approx 69$~mas and exhibits a dust temperature 
of 1000~K. Moreover, Bl\"ocker et al.\ found that in addition to the steady stellar wind 
a phase of enhanced mass loss (superwind) that ceased roughly 60 to 90 yrs ago is required to explain 
both the $K$-band visibility and the SED.
In the intensity profiles shown in Fig.~\ref{fig4_foveffect}b, this temporal superwind phase results in the
bump seen at an angular displacement of $\sim 170$~mas. Thus, according to Bl\"ocker et al.\ \irc's dust shell 
comprises two components: An outer
shell resulting from the previous superwind phase and, directly adjacent to this outer shell, 
an inner shell resulting from the present-day mass loss which followed the superwind phase.
During the superwind phase, the mass-loss rate was approximately 40
times higher than the current mass-loss rate of $1.4\times 10^{-5}\,M_\odot{\rm yr}^{-1}$.

%%%%%%%%%%%%%%%%%%%%%%%%%%%%%%%%%%%%%%%%%%%%%%%%%%%%%%%%%%%%%%%%%%%%%%%%%%%%%%%%%%%%%%%%%%%%%%%%%%%%%%%%%%
%%%
%%% Fig.5: AMBER LR observables -rescaled
%%%
%%%%%%%%%%%%%%%%%%%%%%%%%%%%%%%%%%%%%%%%%%%%%%%%%%%%%%%%%%%%%%%%%%%%%%%%%%%%%%%%%%%%%%%%%%%%%%%%%%%%%%%%%%
   \begin{figure*}
	\vspace*{-3mm}

   \centering
   \includegraphics[angle=0,width=0.8\textwidth]{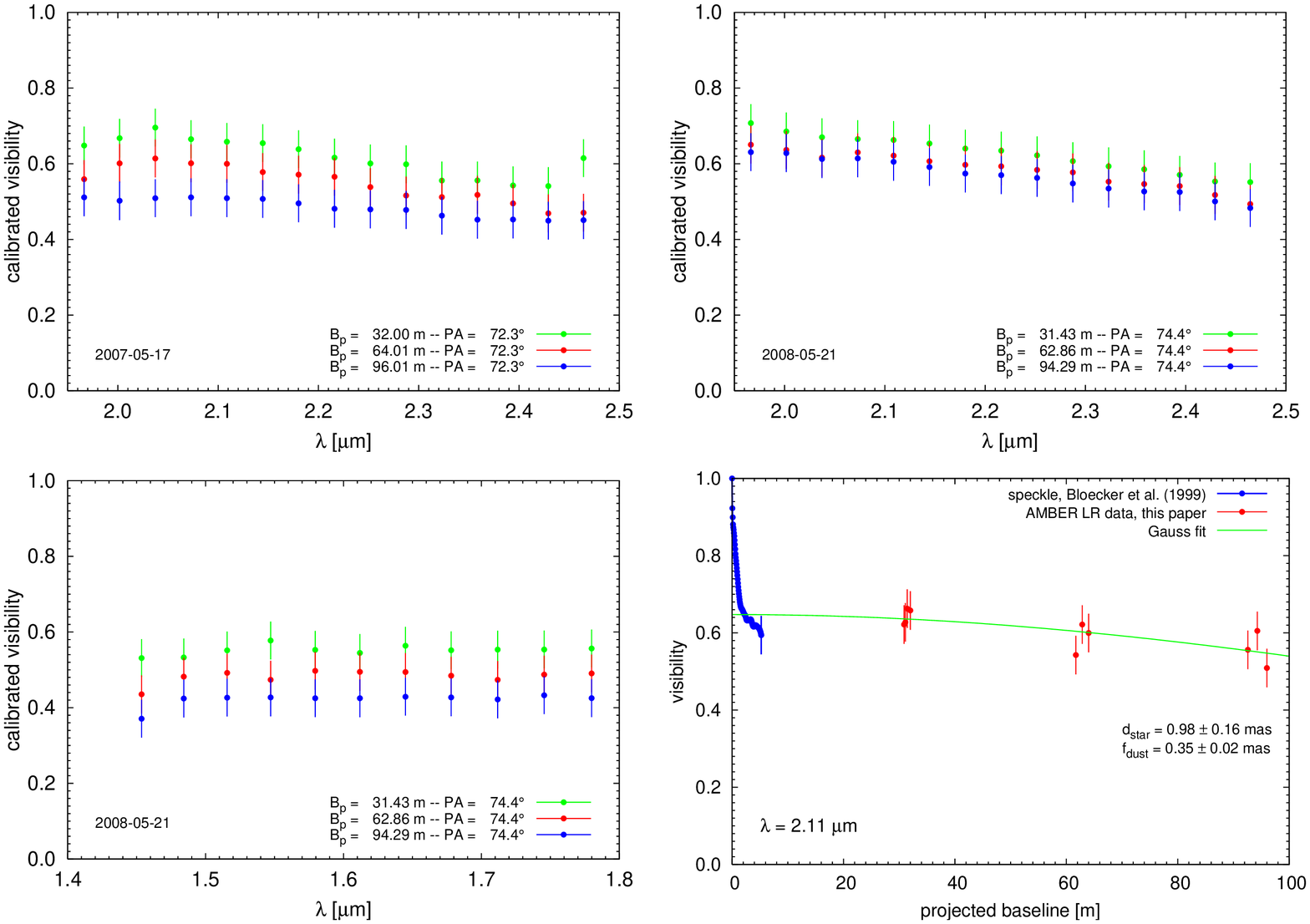}
      \caption{VLTI/AMBER observations of \irc corrected for the field-of-view effect. 
       {\it Top:} The two panels show the $K$-band AMBER visibilities as a function of wavelength for two of our 
                  LR measurements in May 2007 and May 2008, corrected for the limited FOV, as discussed in 
                  Sect.~\ref{sect_res_fov}. 
       {\it Bottom left:}
                  AMBER $H$-band visibilities as a function of wavelength from our LR observations in May 2008,
                  corrected for the limited FOV.
       {\it Bottom right:}                  
		  AMBER $K$-band visibilities at $\lambda=2.11\,\mu$m corrected for AMBER's limited FOV 
                  (red bullets) and visibility from the speckle-interferometric observations of 
                  \citet[blue bullets]{bloecker99} as a function of baseline. As the figure reveals, 
		  after correction for the FOV effect the speckle and long-baseline observations are in good
                  agreement, showing the fully resolved dust shell and the nearly unresolved stellar component.
                  The solid green line is a two-component fit of a compact Gaussian representing the stellar
                  component and a fully resolved dust shell.
                  The fit reveals a Gaussian FWM diameter of $d_\star=0.98\pm0.16$~mas and a flux contribution 
                  of $f_{\rm dust}=0.35\pm0.02$ from the dust shell.
              }
         \label{fig5_lr_data_rescaled}
	\vspace*{-4mm}

   \end{figure*}
%%%%%%%%%%%%%%%%%%%%%%%%%%%%%%%%%%%%%%%%%%%%%%%%%%%%%%%%%%%%%%%%%%%%%%%%%%%%%%%%%%%%%%%%%%%%%%%%%%%%%%%%%%

Since the speckle data of Bl\"ocker et al.\ were taken with a comparably large FOV 
($\sim14$\arcsec), the infinite FOV of the DUSTY model was well suited for the speckle data analysis. 
In the context of our AMBER observations, we re-examined the final radiative transfer model of Bl\"ocker 
et al.\ as follows. We used the radial $H$- and $K$-band intensity profiles of their final 
model (see the top-right panel in Fig.~\ref{fig4_foveffect}) and calculated the visibilities from the intensity
profiles, which were truncated to match a given FOV. We note that it would have been more precise to convolve
the intensity profiles with the corresponding telescope aperture before the truncation, but the effect of the
convolution on the final result is only weak and, therefore, has been omitted here. 
The $H$- and $K$-band visibilities resulting from our FOV correction are shown 
in the middle panels of Fig.~\ref{fig4_foveffect} for the FOVs given by the plot labels. As expected,
the smaller the FOV, the higher the plateau value of the visibility, since an increasing fraction of the
extended dust shell lies outside the FOV and will, therefore, no longer contribute to the flux and visibility,
respectively. 

In the bottom right panel of Fig.~\ref{fig4_foveffect}, the AMBER visibility correction for \irc is shown 
as a function of wavelength across the $H$ and $K$ bands. This wavelength-dependent correction has to be
subtracted from the measured visibility in order to correct AMBER's finite FOV when observing with
the ATs. The figure reveals that the visibility correction decreases with 
increasing wavelength from approximately 0.42 at $1.45\,\mu$m to 0.1 at $2.45\,\mu$m.
This decrease reflects the decreasing flux contribution of the dust shell (scattering + thermal emission) 
with increasing wavelength (see upper right panel in Fig.~\ref{fig4_foveffect}). For instance, at $1.5\,\mu$m
the contribution of the attenuated stellar flux to the stellar flux is only 35\%, while it is almost
60\% at $2.2\,\mu$m. Correspondingly, the FOV correction is stronger in the $H$ band than in the $K$ band.
At the center of the $H$ and $K$ bands the visibilities of the unresolved stellar component in the 
model with a 250~mas FOV are approximately higher by 0.18 and 0.35 compared to the DUSTY model with an 
infinite FOV. For a 60~mas FOV, the effect is even more dramatic. Here, almost the complete dust shell is 
truncated due to the small FOV, and the visibility approaches unity, since mainly the compact stellar 
component contributes to the total flux. According to Fig.~\ref{fig4_foveffect}, we can expect that only 
$\sim 6$\% instead of the originally 40\% of flux from the extended dust shell remains, if the FOV is as 
small as 60~mas.

%%%%%%%%%%%%%%%%%%%%%%%%%%%%%%%%%%%%%%%%%%%%%%%%%%%%%%%%%%%%%%%%%%%%%%%%%
\subsection{The low-spectral resolution data}\label{sect_res_lr}
%%%%%%%%%%%%%%%%%%%%%%%%%%%%%%%%%%%%%%%%%%%%%%%%%%%%%%%%%%%%%%%%%%%%%%%%%

As discussed in the previous section, the AMBER visibilities of \irc obtained with both ATs and
UTs are highly affected by the limited FOV of the observations. From the radiative transfer model of \cite{bloecker99},
we estimated the effect of AMBER's limited FOV on the $H$- and $K$-band visibilities for baselines which start to
resolve the stellar component and its dense wind. We lowered all LR $H$- and $K$-band visibilities shown in 
Fig.~\ref{fig2_lrdata} according to the wavelength-dependent visibility correction shown in the bottom panel of
Fig.~\ref{fig4_foveffect}. Figure 
\ref{fig5_lr_data_rescaled} shows the rescaled $K$-band AMBER visibilities from May 2007 and May 2008
(top panels) and the rescaled $H$-band AMBER visibilities from May 2008 (bottom left panel) which would be obtained
in the case of an infinite AMBER FOV. 

As the bottom-right panel in Fig.~\ref{fig5_lr_data_rescaled} illustrates,
the rescaled AMBER $K$-band data at $\lambda=2.11\,\mu$m (red bullets with error bars) are in good agreement with 
the visibilities from the speckle-interferometric observations by \cite{bloecker99} (blue bullets) within the 
uncertainties of both measurements. The solid green curve in
Fig.~\ref{fig5_lr_data_rescaled} shows a Gaussian fit of the compact stellar component, assuming a fully resolved dust
shell with an a priori unknown flux contribution. As indicated in the plot, a two-parameter fit with stellar diameter
$d_{\star}$ and the fractional flux contribution of the dust shell as $f_{\rm dust}$ free parameters gives a Gaussian 
FWHM diameter of $d_\star=0.98\pm0.16$~mas and a dust shell flux contribution of $f_{\rm dust}=0.35\pm0.02$. 
The fitted diameter is in agreement with the stellar diameter $d=1.01$~mas derived for a bolometric flux of
$8.2\times~10^{-10}\,{\rm Wm}^{-2}$, a distance of 5~kpc, and a central star effective temperature of 7000~K 
\citep[see, e.g.,][]{bloecker99}.

%%%%%%%%%%%%%%%%%%%%%%%%%%%%%%%%%%%%%%%%%%%%%%%%%%%%%%%%%%%
%%%
%%% Fig.6: diameter vs. wavelength for the LR observations
%%%
%%%%%%%%%%%%%%%%%%%%%%%%%%%%%%%%%%%%%%%%%%%%%%%%%%%%%%%%%%%
   \begin{figure}[h]
	\vspace*{-3mm}

   \centering
   \includegraphics[angle=-90,width=0.4\textwidth]{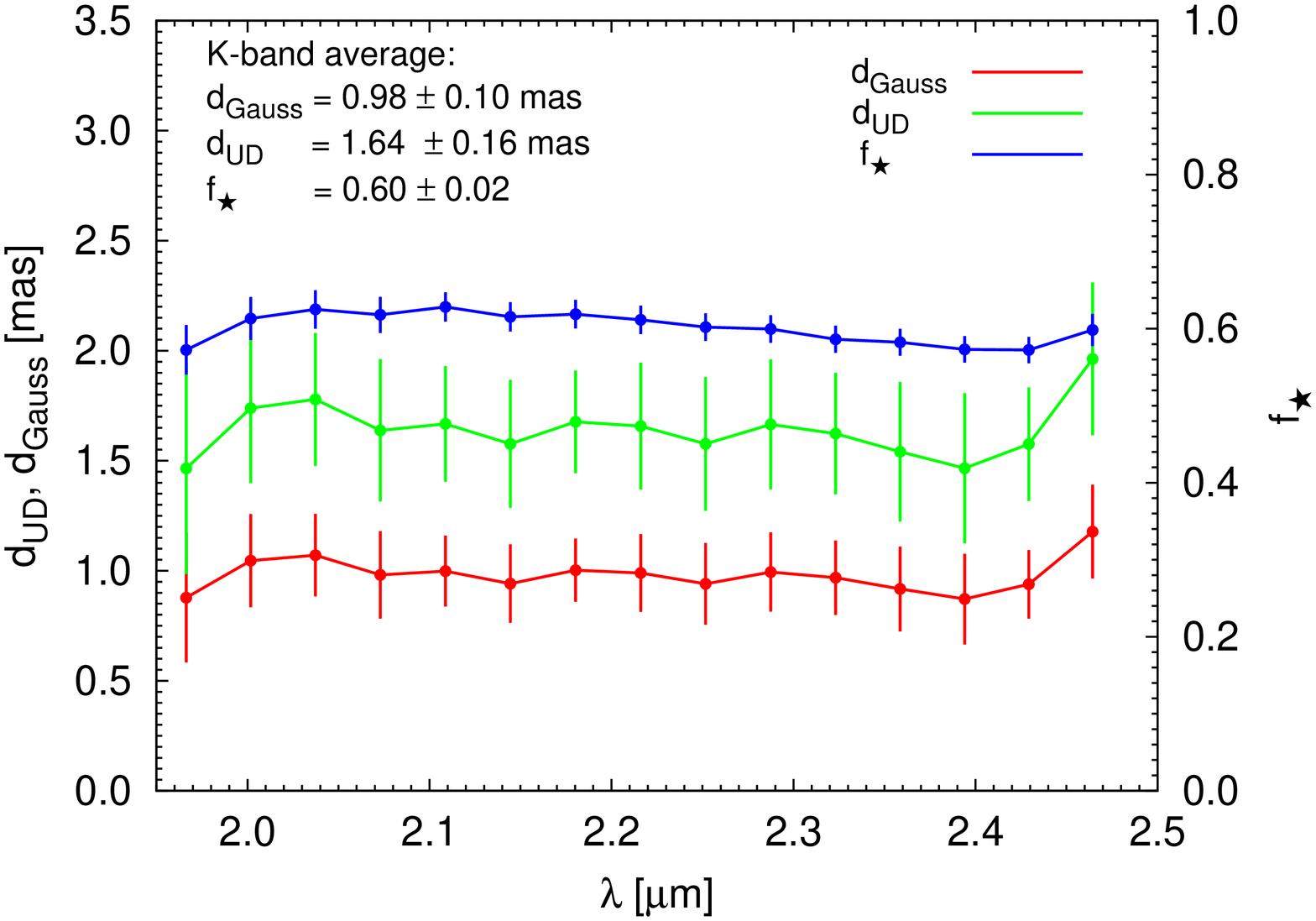}\\
   \includegraphics[angle=-90,width=0.4\textwidth]{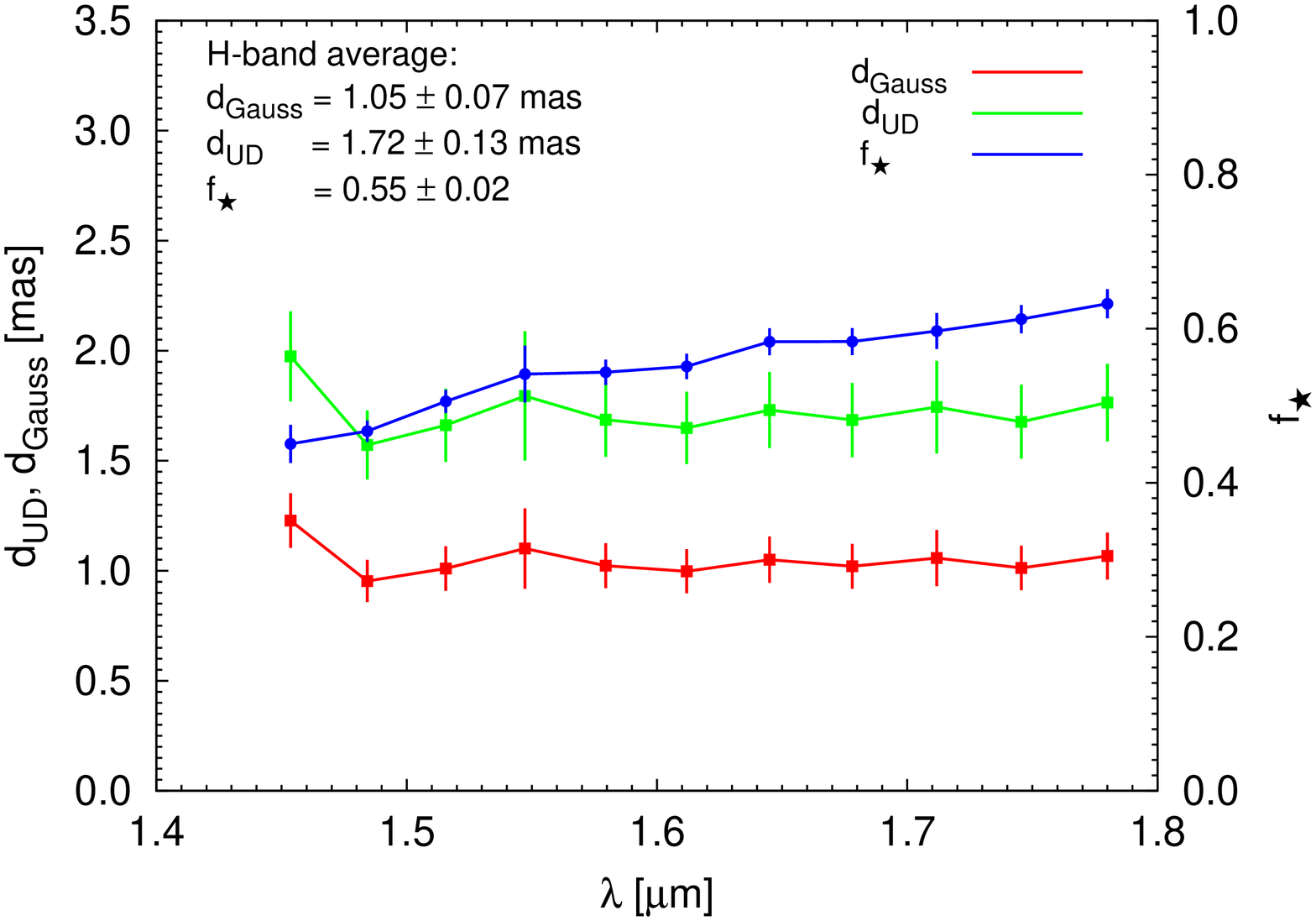}
      \caption{Stellar diameter $d_\star$ (red: Gaussian; green: uniform disk) and fractional stellar flux contribution 
       $f_\star$ (blue curve; right y-axis scale) as a function of wavelength resulting from two-component fits 
       (star + very extended dust shell) of the AMBER LR data of \irc in the $K$ (top) and $H$ (bottom) bands. 
        Averaged over 
        all wavelength channels, we find a Gaussian FWHM stellar diameter $d_{\rm Gauss} = 0.98\pm0.10$~mas 
        (uniform disk: $d_{\rm UD} = 1.64\pm0.16$~mas) in the $K$ band and $f_\star = 0.60\pm0.02$. From the AMBER
        $H$-band data, averaged over all wavelength channels, we find a diameter of $d_{\rm Gauss} = 1.05\pm0.07$~mas 
        ($d_{\rm UD} = 1.72\pm0.13$~mas) and a fractional flux contribution of $f_{\rm dust} = 0.55\pm0.02$ from the
        dust shell. The derived flux contribution of the central star and its wavelength dependence 
        is in basic agreement with the predictions from the DUSTY model of \citet{bloecker99}. 
              }
         \label{fig6_lrdiameters}
	\vspace*{-5mm}

   \end{figure}
%%%%%%%%%%%%%%%%%%%%%%%%%%%%%%%

%%%%%%%%%%%%%%%%%%%%%%%%%%%%%%%%%%%%%%%%%%%%%%%%%%%%%%%%%%%%%%%%%%%%
%%%
%%% Fig.7: rescaled visib. around Brgamma + best fit at line center
%%%
%%%%%%%%%%%%%%%%%%%%%%%%%%%%%%%%%%%%%%%%%%%%%%%%%%%%%%%%%%%%%%%%%%%%
   \begin{figure}[h]
	\vspace*{-3mm}

   \centering
   \includegraphics[angle=-90,width=0.4\textwidth]{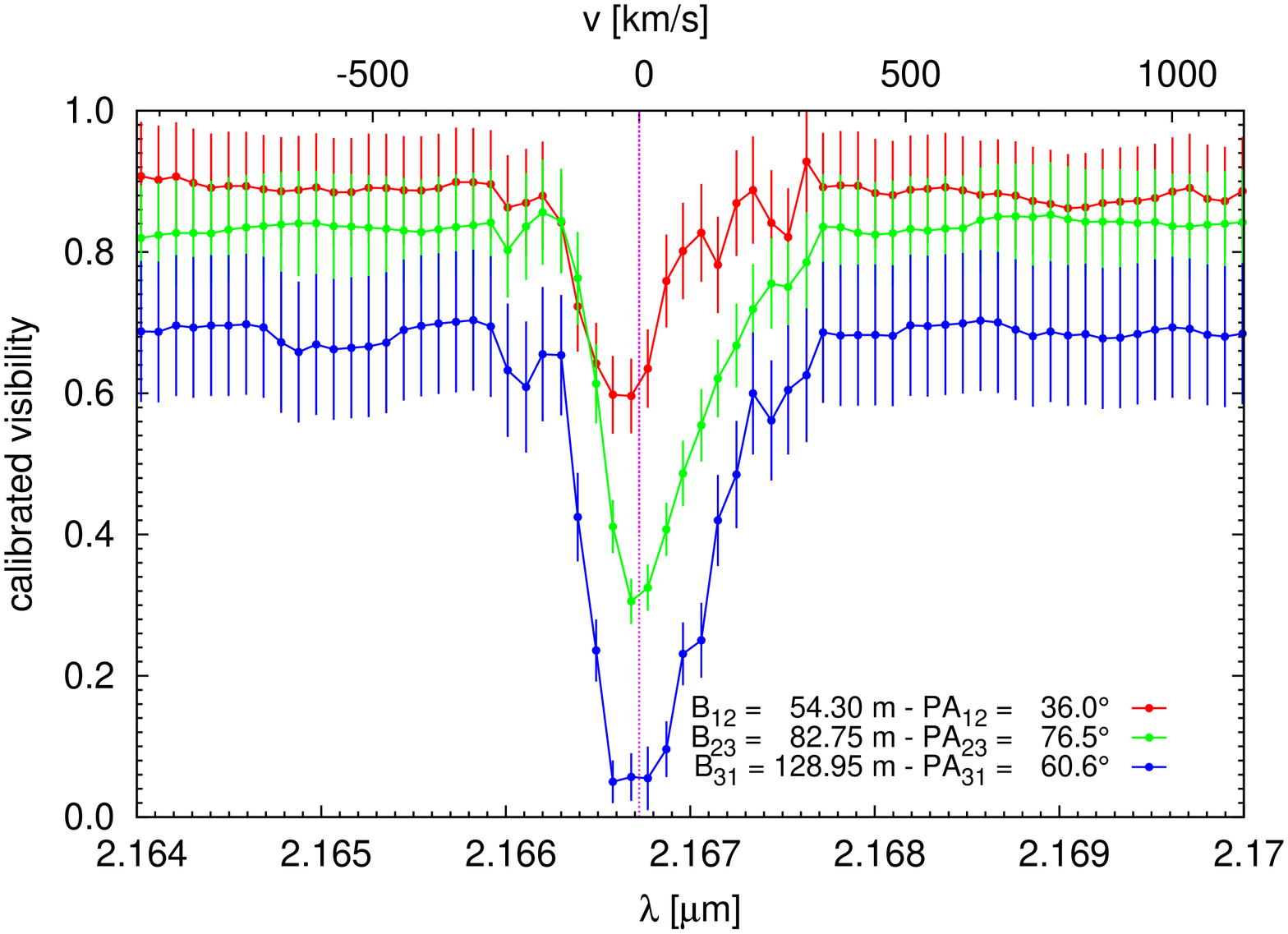}\\
   \includegraphics[angle=-90,width=0.4\textwidth]{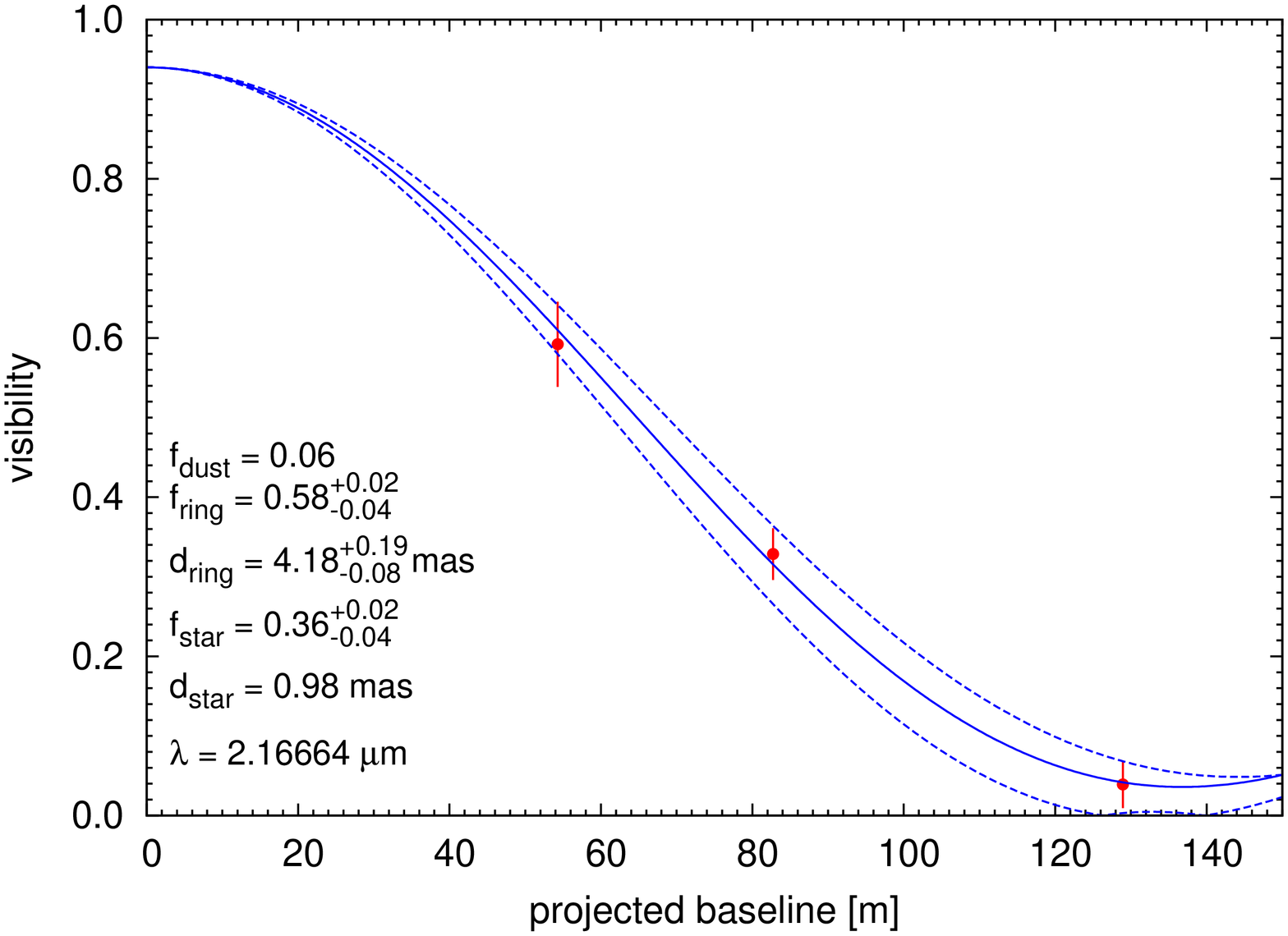} \\
   \includegraphics[angle=-90,width=0.3\textwidth]{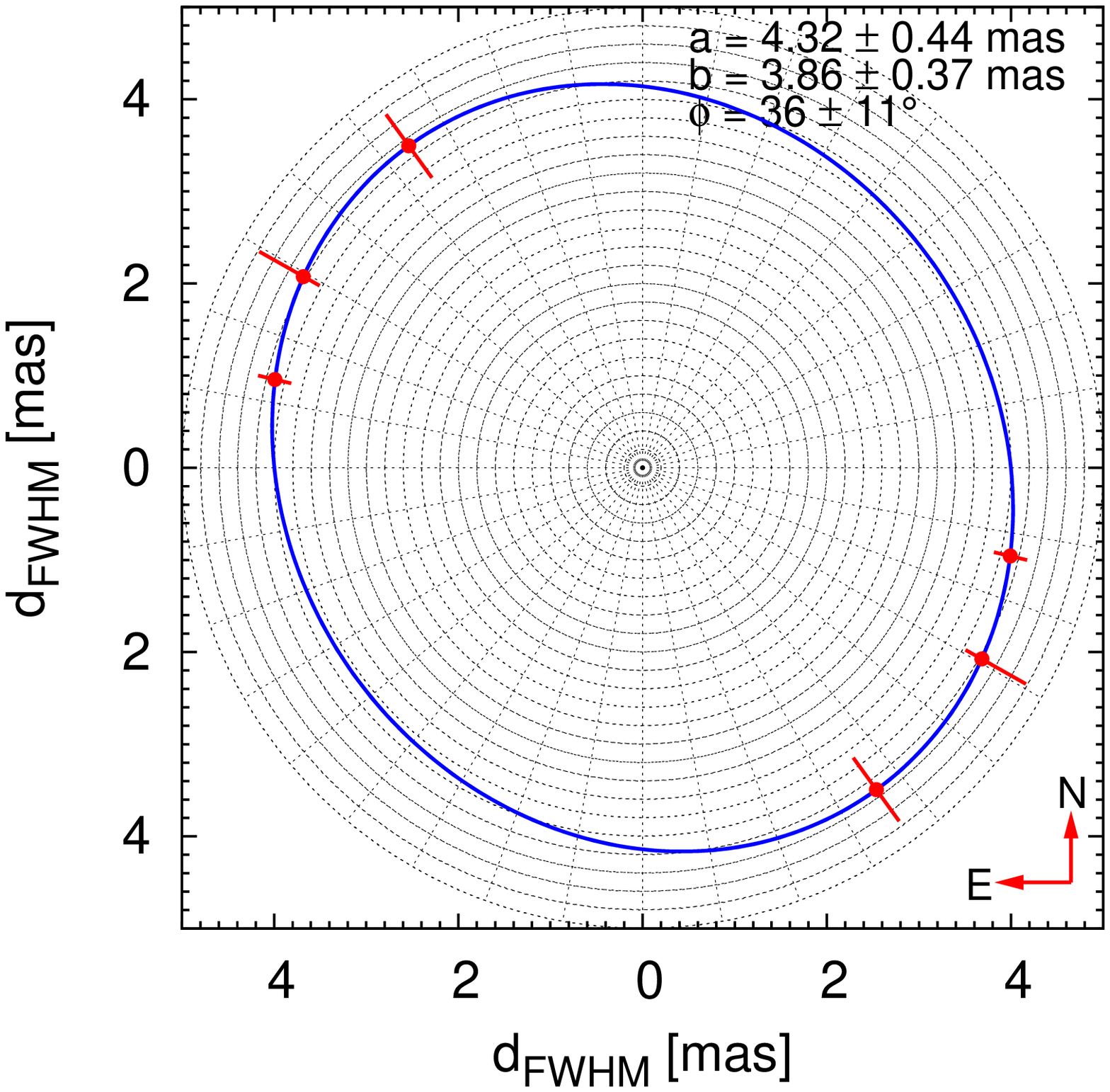}	
      \caption{{\it Top:} HR AMBER visibilities of \irc around the \brg-emission line as a function of wavelength.
               The data have been rescaled according to the procedure described in Sects.~\ref{sect_res_fov} and
               \ref{sect_res_hr} to account for AMBER's 60~mas FOV. The strong decline of the visibilities across
               the \brg line illustrate that the \brg line-emitting region is fully resolved by our observations.
		{\it Middle:}
		AMBER HR visibilities for $\lambda=2.16664\,\mu$m (red bullets with error bars) and
                a corresponding three-component fit (dust shell + stellar continuum + line-emitting region;
                solid blue line with tolerance fits represented by dashed blue lines) 
                according to Eqs.~(\ref{eq1})-(\ref{eq6}). From the fit, we find that the
                diameter of the line-emitting region is 
                $d_{\rm Br\gamma} = 4.18^{+0.19}_{-0.09}\,{\rm mas}\,= 4.12 \times d_{\star}$.
                See Sect.~\ref{sect_res_hr} for further details.
		{\it Bottom:}
		Polar diagram showing Gaussian FWHM ring diameters fitted to the three AMBER visibilities 
                at $\lambda=2.16664\,\mu$m (red bullets with error bars) together with an ellipse fit (blue
                solid line). As the figure illustrates, the fit supports the idea that the region of the
                \brg emission is elongated towards a position angle of $36\pm11$\degr, but due to the very 
                limited amount of data, this result might be biased and has to be taken with care.
              }	
         \label{fig7_hrrescaledfit}
	\vspace*{-4mm}

   \end{figure}
%%%%%%%%%%%%%%%%%%%%%%%%%%%%%%%%%%%%%%%%%%%%%%%%%%%%%%%%%%%%%%%%%%%%%%%%%%%%%%%%%%%%%%%%%%%%%

After rescaling all AMBER $H$- and $K$-band visibilities, we fitted all visibility points for a given wavelength
with the simple two-component model described above (Gaussian stellar component + infinitely extended dust shell).
The result of the fit procedure is shown in Fig.~\ref{fig6_lrdiameters}, where the fit parameters are displayed as a 
function of wavelength for the $H$- (bottom panel) and $K$-band data (top panel). As the figure reveals, within the 
error bars we essentially obtained both a
wavelength-independent continuum diameter and a wavelength-independent flux contribution from the extended 
dust shell across the $K$ band. On the other hand, a slight decrease of the dust shell's flux contribution 
is seen across the $H$ band from $f_{\rm dust}\approx 0.64$ at the lower band edge to $f_{\rm dust}\approx 0.54$ at 
the upper band edge. A comparison with the lower left panel in Fig.~\ref{fig4_foveffect} shows that the wavelength dependence
and the absolute level of the stellar flux contribution derived from the AMBER data is in basic agreement with the predictions
from the DUSTY model of \citet{bloecker99}. In this model, the attenuated stellar flux amount to approximately 0.58 in the
$K$ band with only a moderate wavelength dependence, while the stellar flux contribution rises from approximately
0.32 to 0.5 across the $H$ band.

Averaged over all $K$-band data, for \irc we derived a stellar continuum diameter of
$d_{\rm Gauss} = 0.98\pm0.10$~mas ($d_{\rm UD} = 1.65\pm0.14$~mas assuming a uniform disk) and a fractional flux
contribution of $1-f_{\rm\star} = f_{\rm dust} = 0.41\pm0.05$ from the dust shell. From the $H$-band AMBER data, 
we found $d_{\rm Gauss} = 0.99\pm0.07$~mas ($d_{\rm UD} = 1.63\pm0.12$~mas) and $f_{\rm dust} = 0.44\pm0.01$.

%%%%%%%%%%%%%%%%%%%%%%%%%%%%%%%%%%%%%%%%%%%%%%%%%%%%%%%%%%%%%%%%%%%%%%%%%
\subsection{The high-spectral resolution data}\label{sect_res_hr}
%%%%%%%%%%%%%%%%%%%%%%%%%%%%%%%%%%%%%%%%%%%%%%%%%%%%%%%%%%%%%%%%%%%%%%%%%

%%%%%%%%%%%%%%%%%%%%%%%%%%%%%%%%%%%%%%%%%%%%%%%%%%%%%%%%%%%%%%%%%%%%%%%%%
\subsubsection{FOV effect correction}\label{sect_res_hr_fov}
%%%%%%%%%%%%%%%%%%%%%%%%%%%%%%%%%%%%%%%%%%%%%%%%%%%%%%%%%%%%%%%%%%%%%%%%%

To account for the larger uncertainties in the absolutely calibrated data, we used a slightly different strategy
to correct the HR AMBER data for the FOV effect described in Sect.~\ref{sect_res_fov}. Looking 
at Fig.~\ref{fig4_foveffect}e we see that for AMBER+UT observations with a FOV of $\sim60$~mas, the fractional flux
contribution of the dust shell in the $K$ band is only of the order of 6\%. On the other hand, from the fit
of our LR AMBER data, we found that \ircend's $K$-band continuum diameter is 0.98~mas. Therefore, we corrected
our HR AMBER visibilities in the following way:
For a given baseline, we first rescaled the visibility as a function of wavelength to match the value expected
for a single, compact component with 0.98~mas diameter. Then, we globally lowered the rescaled visibilities 
by 6\% to account for the diffuse, extended flux contribution of the dust shell.

The final result of this rescaling process in order to take the FOV effect into account is shown in the upper 
panel of Fig.~\ref{fig7_hrrescaledfit}. Here, we show for a small wavelength range around the \brg line the 
$K$-band visibilities one would retrieve for \irc in the case of an infinite AMBER FOV. As the figure reveals,
the continuum visibilities are now $\sim0.9$ for the shortest baseline, 0.83 for the intermediate baseline, and
0.68 for the longest baseline. 

%%%%%%%%%%%%%%%%%%%%%%%%%%%%%%%%%%%%%%%%%%%%%%%%%%%%%%%%%%%%%%%%%%%%%%%%%
\subsubsection{Simple geometrical model}\label{sect_res_hr_mod}
%%%%%%%%%%%%%%%%%%%%%%%%%%%%%%%%%%%%%%%%%%%%%%%%%%%%%%%%%%%%%%%%%%%%%%%%%

Attempting to get a first idea of the size of the Br$\gamma$ line-emitting region and
disregarding a possible deviation from spherical symmetry, 
we fitted a spherical Gaussian to all continuum data for a given wavelength with $\lambda < 2.1660\,\mu$m and 
$\lambda < 2.1667\,\mu$m, taking into account the 6\% flux contribution from the extended dust shell. 
Averaged over all spectral channels (approximately 490 channels), we obtained a FWHM Gaussian size of 
$d_{\rm continuum} = 1.014\pm0.01$~mas for the continuum-emitting region. 
We note that the error given here includes not only
the the pure fitting error, which is only of the order of 0.001~mas, but also reflects the uncertainties of
the overall calibration process, including the FOV-related rescaling of the data.

To derive the size of the \brg line-emitting zone, we fitted the visibilities in the spectral channels
with noticeable \brg line-emission in the spectrum, i.e.\ for $2.1660\,\mu{\rm m}\la\lambda\la2.1667\,\mu$m,
with a three-component model. For the sake of simplicity, in this simple model, we neglect likely deviations 
from a spherical or even point-symmetric distribution of the \brg line-emitting region. 
\citet{dewit08} convincingly argued that the \brg line-emitting region cannot have a spherical shape, 
but is most likely elongated towards a position angle of 20\degr. Unfortunately, as in the case of the study
of \citet{dewit08}, also our single measurement with a position angle coverage of $\sim40$\degr\, 
is only of limited use to put reliable constraints on the shape of the emission zone. Nevertheless, we studied
the emission region also in the context of a 2-D radiative transfer model as will be shown in more detail in 
the next section. However, independent from a more sophisticated modeling,
qualitatively we can conclude that the clear non-zero detection in the differential and closure phases 
across the \brg line in our AMBER HR measurement shows that the geometry of the line-emitting region deviates
from a point-symmtric configuration. On the other hand, in the center of the \brg line the differential and 
closure phases are of the order of $\la10$\degr. Thus, in order to determine the typical size of the \brg-emitting zone
with the simple model approach discussed here we decided to neglect the non-zero phases.
 
Our simple three-component model consists of the following components:
(a) a fully resolved, infinitely large dust shell with a fixed fractional flux 
contribution of 6\%, (b) the stellar continuum, approximated by a Gaussian with a fixed size of 0.98~mas 
according to our continuum fit, and (c) the \brg line-emitting region, which is represented in our simple 
model by a ring with infinitely thin thickness ($\Delta r = 0$). The total visibility can then be written as

%%%%%%%%%%%%%%%%
\begin{equation}\label{eq1}
V_{\rm total} = f_{\rm dust} \times V_{\rm dust} + f_{\rm cont} \times V_{\rm cont} + f_{\rm ring} \times V_{\rm ring} 
\end{equation}
with
\begin{equation}
f_{\rm dust} + f_{\rm cont} + f_{\rm ring} = 1.0 \quad {\rm and} \quad f_{\rm dust} = 0.06 \quad {(\rm fixed)},
\end{equation}
\begin{equation}
V_{\rm dust} = 1 \quad {\rm for} \quad B_{\rm p} = 0 \quad {\rm and} \quad V_{\rm dust} = 0 \quad {\rm elsewhere},
\end{equation}
\begin{equation}
%V_{\rm cont} = \exp \{-(\pi d_{\rm cont}B_{\rm p}/\lambda)^2/(4.\log(2.))\}
V_{\rm cont} = \exp \left\{ -\frac{\left(\pi d_{\rm cont}B_{\rm p}/\lambda\right)^2}{4.\times\log(2.)}\right\}
\end{equation}
with $d_{\rm cont} = 0.98$~mas (fixed) and
\begin{equation}\label{eq6}
V_{\rm ring} = J_0(\pi d_{\rm ring}B_{\rm p}/\lambda).
\end{equation}
%%%%%%%%%%%%%%%%

$B_{\rm p}$ is the projected baseline, $J_0$ is the Bessel function of 0th order, and $d_{\rm ring}$ 
is the diameter of the infinitely thin ring representing the \brg line-emitting region.
As can be seen from the above equations, our fit procedure contains two free parameters: the ring
diameter $d_{\rm ring}$ and the fractional flux contribution from the \brg line-emitting region of 
\ircend's optically thick stellar wind, $f_{\rm ring}$. 
In principle, one could also fix $f_{\rm ring}$ and $f_{\star} = 1 - f_{\rm ring} - f_{\rm dust}$ 
according to the spectrum shown in Fig.~\ref{fig3_hrdata}, if the continuum emission
is interpolated across the \brg line. For instance, a linear interpolation from the spectral 
channels measuring pure continuum emission would give $f_{\star} = 0.415$ and $f_{\rm ring} = 0.525$.
But since it is not clear a priori to which extent photospheric absorption is present \citep[which is probably
not unlikely for an A-type star, see discussion in][]{dewit08}, we decided to keep the fractional flux 
contribution from the stellar wind component as a free parameter.

Our best-fit result for the central wavelength channel in the \brg line is shown in the middle panel of 
Fig.~\ref{fig7_hrrescaledfit}. Here, the visibility according to Eqs.~(\ref{eq1})-(\ref{eq6}) is plotted 
as a function of baseline for $\lambda=2.16664\,\mu$m. The blue bullets with error bars are
the three AMBER data points, and the three curves visualize the best fit model and its tolerance. We found
that our AMBER data in the center of the \brg emission line can be well fitted with a ring with a diameter of
$d_{\rm ring} = 4.18^{+0.19}_{-0.09}~{\rm mass}\,\approx 4.12 \times d_{\star}$, whose fractional flux 
contribution is $f_{\rm ring} = 0.58^{+0.02}_{-0.04}$. We note that the size of the \brg line-emitting 
region obtained from our fit is in general agreement, although slightly larger than the 3.3~mas 
found by \citet{dewit08} from a single-component Gaussian fit of the line data.
From $f_{\rm ring} $, we directly obtain $f_{\star} = 0.36^{+0.02}_{-0.04}$;
i.e., from our best fit, we indeed find a photosperic absorption in the \brg line of the order of
$f_{\rm abs} = (f_{\star, {\rm cont}} - f_{\star, {\rm line}})/f_{\star, {\rm cont}} = 
(0.415 - 0.36)/0.415 = 0.13$.

Despite the limited amount of data and the limited position angle coverage, we also briefly addressed the
question of an aspherical shape of the \brg line-emitting region in the context of our simple model fits. 
For this study, we used the three-component 
model described above and fixed $f_{\rm ring}$ to the value found from the 1-D fit ($f_{\rm ring} = 0.58$)
of the $\lambda=2.16664\,\mu$m data. Then, we first fitted all three data points with the three-component model
with only $d_{\rm ring}$ as free parameter, and finally, we fitted an ellipse to the three ring diameters as
a function of position angle. The result is shown in the bottom panel of Fig.~\ref{fig7_hrrescaledfit}, where
the fitted ring diameters and the ellipse are displayed in a polar diagram. As the figure shows, the position-angle
dependence of the ring diameter can be fitted with an ellipse with major and minor axes of $4.32\pm0.44$~mas
and $3.86\pm0.37$~mas, respectively, and a position angle of the major axis of $\Phi=36\pm11$\degr. Interestingly,
the major axis is nearly aligned with the elongation of the outer reflection nebula of \irc seen
in the HST images of \citet[][PA = 33\degr]{humphreys97} and the symmetry axis of the H$\alpha$ emission 
as inferred by \citet{davies07}. 

%%%%%%%%%%%%%%%%%%%%%%%%%%%%%%%%%%%%%%%%%%%%%%%%%%%%%%%%%%%%%%%%%%%%%%%%%%%%%%%%%%%%%%%%%%%%
\subsubsection{Radiative transfer modeling of the Br$\gamma$ line emission}\label{radtransf}
%%%%%%%%%%%%%%%%%%%%%%%%%%%%%%%%%%%%%%%%%%%%%%%%%%%%%%%%%%%%%%%%%%%%%%%%%%%%%%%%%%%%%%%%%%%%

A simple geometrical model as presented in the previous section cannot explain the AMBER observations in detail.
Thus, a more physical model is required to interpret, for instance, the blueshift of the Br$\gamma$ emission,
the visibility across the emission line, and the non-zero phases. Therefore, in a second step of our analysis
we interpreted the AMBER data of \irc using the iterative, spherical symmetric, full line-blanketed, non-LTE 
radiative transfer code CMFGEN \citep{hm98}. 

%%%%%%%%%%%%%%%%%%%%%%%%%%%%%%%%%%%%%%%%%
\paragraph{The radiative transfer code}
%%%%%%%%%%%%%%%%%%%%%%%%%%%%%%%%%%%%%%%%%

In CMFGEN, a spherically-symmetric outflow in steady-state is assumed, and line and continuum formation are
calculated in a non-LTE regime. Each model is specified by its effective temperature \teff, luminosity \lstar, 
effective gravity \geff, mass-loss rate \mdot, wind terminal velocity \vinf, and chemical abundances $Z_i$ 
of the included species. Moreover, a velocity law must be adopted, since the momentum equation of the wind is not solved. 
CMFGEN can only handle monotonically 
increasing velocity laws and, therefore, no inflows can be analyzed using this code. The velocity structure $v(r)$ is 
parameterized by a beta-type law, which is modified at depth to smoothly match a hydrostatic structure at $\rstar$. 
We define the hydrostatic radius as the radius where the wind velocity is equal to one third of the sonic velocity 
($v=v_\mathrm{sonic}/3$), in order to avoid any effect of the dense wind on the determination of $\rstar$. Close to and 
below $\rstar$, the hydrostatic equation is iterated in order to produce a quasi-hydrostatic structure which extends 
inwards to the inner model boundary (at a Rosseland optical depth of 100). The effective temperature is defined 
as the temperature where the Rosseland optical depth is 2/3, i.e., $\teff=T(\tau_\mathrm{Ross}=2/3$).

While CMFGEN can handle the effects of clumping via a volume filling factor $f$, we decided to assume a unclumped wind 
(i.e.\ $f=1$) since no strong electron scattering wings are present around the \brg line. 
The strength of such electron scattering wings is sensitive to the amount of wind clumping \citep{hillier91}, 
and even assuming an unclumped wind model (which produces the maximum strength possible in the electron scattering wings), 
no detectable electron scattering wings are seen in \brg. 

Line blanketing affects the ionization structure of the wind and, consequently, the spectrum. CMFGEN employs the 
concept of super-levels to include thousands of spectral lines in non-LTE, making it feasible to solve the 
equations of statistical equilibrium and radiative transfer simultaneously. The atomic model used for \irc included lines 
of H, He, C, N, O, and Fe.

Although CMFGEN handles only spherical-symmetric outflows, the wind asymmetry was also analyzed using a recently developed 
modification in CMFGEN to compute the emerging spectrum in two-dimensional geometry \citep{bh05}. We refer the reader 
to \citet{bh05} for further details about the code, and to that paper and \citet{ghd06,groh08} for additional details about 
applications of the BH05 code. In the following, we briefly describe the main aspects of the code.

As input the BH05 code requires several quantities (energy-level populations, ionization structure, temperature structure, 
emissivities, opacities, and specific intensity $J$) calculated by the original, spherically-symmetric CMFGEN model in the 
co-moving frame. Latitude-dependencies of the wind density and wind terminal velocity can then be taken into account using 
any arbitrary latitude-dependent density/wind terminal velocity variation. Due to our very limited amount of data, we restricted 
the analysis to changes only in the wind density, using oblate and prolate density parameterizations as follows,

%%%%%%%%%%%%%%%%
\begin{equation}
\mathrm{oblate}: \,\,\, \rho_{2D} \propto \rho_{1D} \{1+ [a (1-cos^2 \theta)^b]\}\,\,
\end{equation}
\begin{equation}
\mathrm{prolate}: \,\,\, \rho_{2D} \propto \rho_{1D} (1 + a . cos^b \theta)\,\,,
\end{equation}
%%%%%%%%%%%%%%%

where $\theta$ is the latitude angle (0$^{\circ}$=pole, 90$^{\circ}$=equator). A scaling law is used to ensure that the 
2-D model has the same mass-loss rate as an equivalent spherically symmetric model.

The scaling laws given above are applied only for distances greater than that where the wind velocity is equal to the sonic 
velocity. For smaller distances, no scaling was applied. The 2-D source function, emissivities, and opacities are then 
calculated, assuming 
that these quantities depend only on the new values of the scaled density. Appropriate scaling laws are used for different 
physical processes (e.g.\ density-squared scaling for free-free and bound-free transitions, and linear-density scaling for 
electron scattering). Finally, for a direct comparison with the AMBER observables
the spectrum in the observer's frame 
and intensity maps for all wavelengths of interest are computed. In our analysis, we first considered a spherically symmetric
model and then a model with a prolate/oblate wind structure.

%%%%%%%%%%%%%%%%%%%%%%%%%%%%%%%%%%%%%%%%%%%%%%%%%%%%%%%%%
\paragraph{Spherical model}
%%%%%%%%%%%%%%%%%%%%%%%%%%%%%%%%%%%%%%%%%%%%%%%%%%%%%%%%%

%%%%%%%%%%%%%%%%%%%%%%%%%%%%%%%%%%%
%%%%%%%%%%%%%%%%%%%%%%%%%%%%%%%%%%%
%%%%%%%%%%%%%%%%%%%%%%%%%%%%%%%%%%% Spherical CMFGEN model
%%%%%%%%%%%%%%%%%%%%%%%%%%%%%%%%%%%
%%%%%%%%%%%%%%%%%%%%%%%%%%%%%%%%%%%
    \begin{figure*}[th]
	\vspace*{-0mm}
   \centering
	\hspace*{5mm}
   \includegraphics[angle=0,width=0.92\textwidth]{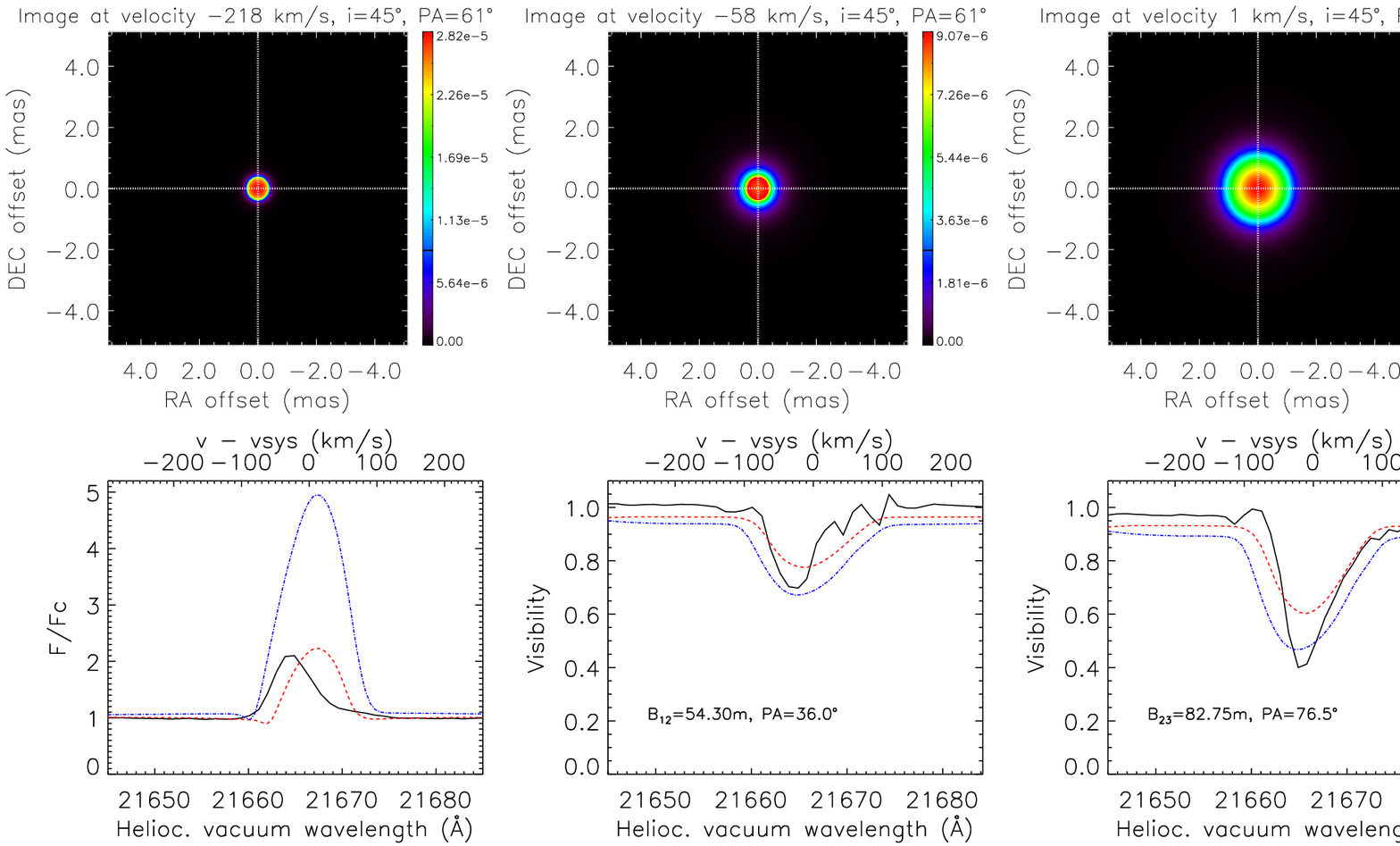}
	\vspace{10mm}
   \caption{{\it Upper row}: Monochromatic images of \irc predicted by the spherical CMFGEN model with the physical parameters 
given in Table \ref{irc_param}. The panels show (from left to right) model images in the $K$ band continuum 
(at a velocity $-218 {\rm kms}^{-1}$, i.e.\ well outside the line), 
the blueshifted \brg emission at $v=-58~\kms$, the center of the \brg emission, and the redshifted \brg emission at $v=41~\kms$. 
The color scale represents the intensity projected on the sky in units of erg/cm$^2$/s/Hz/steradian and is normalized 
to the maximum intensity of each image.
{\it Bottom row, left to right}:  Continuum-normalized AMBER spectrum and visibilities of \irc (black solid line) around \brg 
compared to the spherical CMFGEN model with physical parameters given in Table \ref{irc_param} (red dotted line). 
For comparison, a spherical model with the same parameters but an increased mass-loss rate of 
$\mdot=5 \times 10^{-5}~\msunyr$ is also shown (blue dot-dashed line). 
Note that the observed \brg emission of \irc is blueshifted compared to the spherical models.
Moreover, in disagreement with the observations, spherical models globally predict zero differential and closure phases. 
\label{irc_groh_sph}}
   \end{figure*}
%%%%%%%%%%%%%%%%%%%%%%%%%%%%%%%%%%%%%%%%%%%%%%%%%%%%%%%%%%%%%%%%%%%%%%%%%%%%%%%%%%%

The stellar and wind parameters of our best-fitting spherically symmetric CMFGEN model 
are presented in Table \ref{irc_param}. Our CMFGEN model has a luminosity of $\lstar=6 \times 10^5~\lsun$, in good
agreement with previous determinations \citep[e.g., $\lstar\sim 5 \times 10^5~\lsun$,][]{jones93}
Together with an assumed mass of $M=20~\msun$, we find $\logg=0.8$.
Since \irc has already lost about half of its initial mass \citep{humphreys97}, it is likely that the He 
content is enhanced on the surface compared to the solar value. Thus, apart from solar metallicity 
our model assumes a He abundance of He/H=0.4 (by number). Finally, we assumed a distance of 3.5~kpc. 

The AMBER high-resolution $K$-band spectrum covers only a 
short wavelength region around \brg, and does not provide enough diagnostic lines to 
derive a precise value for the effective temperature of \ircend. However, the absence of \ion{He}{i} $2.16137\mu$m and 
\ion{He}{i} $2.16474\mu$m lines imply an upper limit for $\teff$ of about 10500 K. Since \citet{klochkova02} obtained 
$\teff\simeq9200~\mathrm{K}$ for the observations of \irc in 2000 and an annual increase in $\teff$ of $\sim 120$ K, 
we assume in our models $\teff=9600~\mathrm{K}$. We note that conclusions of the present paper such as the evidence 
for wind asymmetry in \ircend, are only weakly affected by reasonable changes in $\teff$.

Our model requires a mass-loss rate of $\mdot=1.7 \times 10^{-5}~\msunyr$ in order to reproduce the equivalent width 
of the observed Br$\gamma$ line in \irc ($-6.7 {\AA}$, Fig.~\ref{irc_groh_sph}). A wind terminal velocity of 
$\vinf=70~\kms$ and $\beta=1.5$ were used 
in order to roughly reproduce the Br$\gamma$ line width. However, the \brg-emitting region of such a model is rather compact, 
not reproducing the  spatial extension of the \brg emitting region as measured by AMBER. As a consequence, this
model fails to reproduce the observed visibilities (see red curves in the lower panels of Fig.~\ref{irc_groh_sph}). 
Moreover, while the observed spectrum is blueshifted relative to the systemic 
velocity, the CMFGEN model spectrum is slightly redshifted ($\sim 10\kms$). The redshift in the model is caused by the 
turbulent velocity field of the wind \citep{catala84,hillier87a,hillier89}, which was parameterized using a microturbulent 
velocity of $10~\kms$.   

An additional model with the same parameters as the model described above, but with $\mdot=5 \times 10^{-5}~\msunyr$, 
was calculated in order to analyze the effects of a larger $\mdot$ on the fit of the AMBER observables. As expected, 
the increase in $\mdot$ results in a larger Br$\gamma$-emitting region, but while such a model with an increased mass-loss rate 
is now able to reasonably reproduce the observed visibilities (see blue curves in the lower panels of Fig.~\ref{irc_groh_sph}), 
it also predicts a much higher amount of \brg line emission compared to the observations 
(lower left panel in Fig.~\ref{irc_groh_sph}). On the other hand, lower turbulent velocities provide slightly lower 
\brg redshifts. However, as a general result we conclude that spherically symmetric wind models are intrinsically 
unable to reproduce the observed {\it blueshifted} \brg line. Moreover, a reasonable simultaneous fit of both the
observed \brg spectrum and the corresponding visibilities with a spherical model could not be obtained.

%%%%%%%%%%%%%%%%%%%%%%%%%%%%%%%%%%%%%%%%%%%%%%%%%%%%%%%%%%%%%%%%%%%%%%%%%%%%%%%%%%%%%%%%%%%%%%%%%%%%%%%%%%
%%%
%%% Tab.3: CMFGEN Model parameters
%%%
%%%%%%%%%%%%%%%%%%%%%%%%%%%%%%%%%%%%%%%%%%%%%%%%%%%%%%%%%%%%%%%%%%%%%%%%%%%%%%%%%%%%%%%%%%%%%%%%%%%%%%%%%%
\begin{table*}
\caption{Physical parameters of the CMFGEN models for \irc.} 
\label{irc_param}
\centering
\begin{tabular}{l c c c c c c r}
\hline
\hline
\multicolumn{7}{c}{spherical model} \\
\cline{1-7}
\hline
log \lstar/\lsun & \teff  & \logg &  $\mdot$                  & \vinf    & $\beta$ &     He/H    \\
                 &   (K)  &       &  ($\msunyr$)              & ($\kms$) &         &   (number)  \\
\hline
  5.78           &   9600 &  0.8  &  $1.7\times10^{-5}$       & 70       & 1.5     &     0.4 \\
\hline
\\
\hline
\hline
\multicolumn{7}{c}{two-dimensional model} \\
\cline{1-7}
\hline
Density distribution & $a$ & $b$ & i         & PA        & $\mdot$             & \\
                     &     &     & ($\degr$) & ($\degr$) & ($\msunyr$)         & \\ 
\hline
Prolate              &  15 & 2   & 45        & 60        &  $3.5\times10^{-5}$   &\\
\hline
%Oblate & 15 & 2 \\
\hline
\end{tabular}
\end{table*}

%%%%%%%%%%%%%%%%%%%%%%%%%%%%%%%%%%%%%%%%%%%%%%%%%%%%%%%%%%%%%%%%%%%%%%%%%%%%%%%%%%%%%%%%%%%%%%%%%%%%%%%%%%%%%%%%%%%%%%%%%
\paragraph{2-D modeling of the AMBER data around the Br$\gamma$ line - evidence for wind asymmetry}
%%%%%%%%%%%%%%%%%%%%%%%%%%%%%%%%%%%%%%%%%%%%%%%%%%%%%%%%%%%%%%%%%%%%%%%%%%%%%%%%%%%%%%%%%%%%%%%%%%%%%%%%%%%%%%%%%%%%%%%%%

%%%%%%%%%%%%%%%%%%%%%%%%%%%%%%%%%%
%%%%%%%%%%%%%%%%%%%%%%%%%%%%%%%%%%
%%%%%%%%%%%%%%%%%%%%%%%%%%%%%%%%%% 2-D CMFGEN outflow model
%%%%%%%%%%%%%%%%%%%%%%%%%%%%%%%%%%
%%%%%%%%%%%%%%%%%%%%%%%%%%%%%%%%%%
       \begin{figure*}[th]
	\vspace*{0mm}
   \centering
	\hspace*{7mm}
   \includegraphics[angle=0,width=0.94\textwidth]{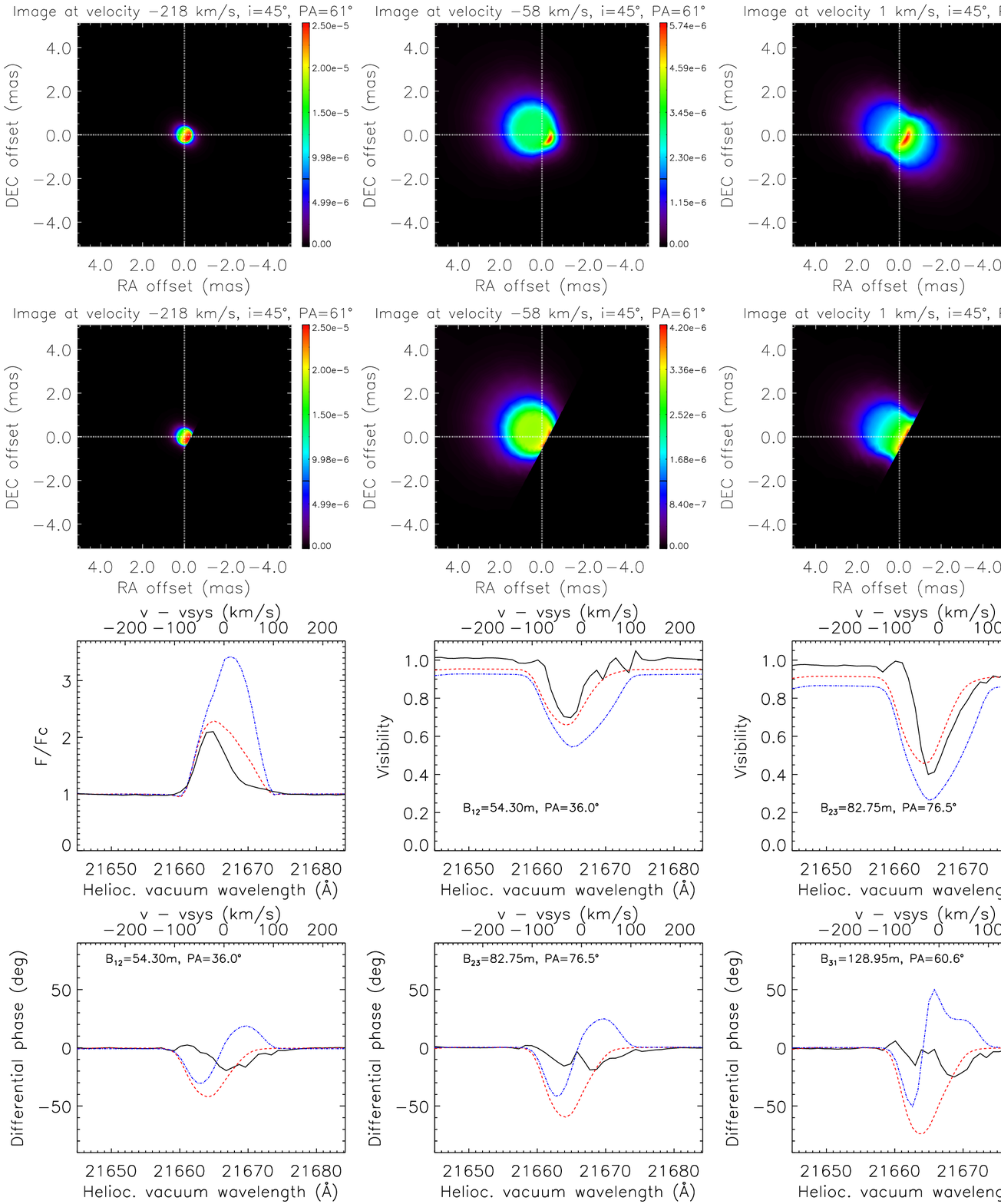}
	\vspace{10mm}
   
   \caption{{\it First row}: Monochromatic images of \irc predicted by the unblocked prolate CMFGEN model with the 
physical parameters given in Table \ref{irc_param}. From left to right, the panels show images in the $K$-band 
continuum (at a velocity $-218 {\rm kms}^{-1}$, i.e.\ well outside the line), 
the blueshifted \brg emission at $v=-58~\kms$, the center of the \brg emission, and the redshifted \brg 
emission at $v=41~\kms$. The color scale represents the intensity projected on the sky in units of 
erg/cm$^2$/s/Hz/steradian and is normalized to the maximum intensity of each image. 
{\it Second row}: Same as the first row, but for a model where the light from the southwestern redshifted lobe blocked. 
For the blocking, a fully opaque screen covering the redshifted lobe and starting at an offset position of 
$\Delta$RA=-0.3~mas and $\Delta$DEC=-0.3~mas was assumed. 
{\it Third row, left to right}:  Continuum-normalized AMBER spectrum and visibilities of \irc (black solid line) around 
\brg compared to a prolate CMFGEN model with physical parameters as given in Table \ref{irc_param} without any blocking 
(blue dot-dashed line) and with the redshift part of the emission blocked (red dotted line). Note that the 
model with the blocked redshifted-lobe reproduces the observed \brg blueshift.
{\it Fourth row, left to right}:  Differential and closure phases of \irc around \brg as observed with AMBER 
(black solid line)  compared to the predictions of the prolate CMFGEN model with physical parameters given in 
Table \ref{irc_param} without any blocking (blue dot-dashed line) and the model where the redshifted \brg emission 
is blocked (red dotted line).\label{irc_groh_2d}}
   \end{figure*} 
%%%%%%%%%%%%%%%%%%%%%%%%%%%%%%%%%%%%%%%%%%%%%%%%%%%%%%%%%%%%%%%%%%%%%%%%%%%%%%%%%%%

In the previous paragraph we showed that spherical models are unable to simultaneously reproduce the observed AMBER 
visibilities and the Br$\gamma$ spectrum of \irc and, in particular, the blueshift of the Br$\gamma$ emission. Thus,
one can conclude that a non-spherical outflow must be present. This is supported by the fact that non-zero differential
and closure phases have been measured with AMBER. Thus, in order to obtain 
insights on the wind asymmetry of \ircend, we calculated 2-D axisymmetric radiative transfer models using the 
\citet{bh05} code described above. 

Due to the limited amount of spatial frequencies and position angles sampled by our data, in our analysis we gave highest 
priority to the simultaneous reproduction of the observed amount of Br$\gamma$ line emission compared to the continuum, 
the blueshift of this emission, and the wavelength dependence of the visibility. Given the limitations of our 2-D model 
and the likely three-dimensional nature of \irc's outflow, a full analysis of the wavelength dependencies of the
differential and closure phases are deferred to a future paper. Nevertheless, we believe that the modeling results  
presented in the following are crucial for constraining the parameter space in future studies using more sophisticated 
2-D radiative transfer models such as ASTAROTH \citep{zsargo08}. 

We analyzed both prolate and oblate density enhancements and found that, in general, both geometries require a remarkably 
large density contrast ($a\sim 7-15$) in order to reproduce the low visibility detected within the Br$\gamma$ line. 
Interestingly, the observed blueshift of the \brg emission could only be reproduced by models in which the receding part 
of the wind, which is responsible for the major fraction of the redshifted Br$\gamma$ emission, is blocked. For a given 
density contrast, oblate models provided less projected separation on the sky between the redshifted and blueshifted 
hemispheres than prolate models, making it challenging for oblate models to produce a blueshifted Br$\gamma$ line with 
a similar strength and spatial extension as the observations. Therefore, the preferred geometry to fit the observations 
is prolate and the blocking of the redshifted hemisphere could possibly be caused either by an optically-thick disk, 
a putative binary companion, or by the wind being ejected by specific parts of the stellar surface. 

The best fit to simultaneously reproduce the observed amount of Br$\gamma$ line emission with respect to the continuum, 
the line blueshift, and the wavelength dependence of the visibility was obtained by using a prolate model with a density 
contrast of $\sim 16$ from pole to equator (see Fig.~\ref{irc_groh_2d}) oriented at $\mathrm{PA}=60 \degr$. The optimal 
fit to the line's blueshift was obtained by starting the blocking of the redshifted emission at a projected offset 
distance from the center of the star of $\Delta$RA=-0.3~mas and $\Delta$DEC=-0.3 mas (Fig.~\ref{irc_groh_2d}).
As Fig.~\ref{irc_groh_2d} reveals, while the 2-D models can basically resemble the AMBER spectrum and visibilities across
the \brg line and, in the the case of the blocked model, also the observed blueshift of the emission, neither the unblocked 
nor the blocked model is able to reproduce the observed wavelength dependence of the differential and closure phases.
Both models predict much stronger phase signals across the line, especially in the blueshifted wing of the emission line.
As noted above, a consistent fit of all AMBER observables including the phases is beyond the scope of this paper and will
be subject to a more detailed analysis which also includes follow-up AMBER oberservations of \irc which will help to
better constrain the 2-D model. 

Compared to our spherical model presented before, the mass-loss rate of the 2-D model had to be increased to 
$\mdot=3.5 \times 10^{-5}~\msunyr$, since a fraction of the \brg emission is blocked. Lower polar density enhancements 
did not provide enough extension of the \brg line-emitting region, while higher mass-loss rates provided a stronger 
\brg emission line than observed with AMBER. The value of $\mathrm{PA}=60\degr$ is admittedly biased by the 
visibility measurement of the longest baseline, which had an orientation on the sky of $60.6\degr$. Different PAs 
provide a worse fit to the \brg visibilities and the differential and closure phases have less amplitude than 
the model with $\mathrm{PA}=60 \degr$. Interferometric measurements using baselines oriented at different PAs on 
the sky are urgently needed to better constrain the position angle of \hbox{\irc's} prolate wind.

An intermediate inclination angle of $i=45 \pm 15 \degr$ is required in order to provide enough separation between 
the projected redshifted and blueshifted lobes and the star. In this case, the redshifted part of the line emission 
can be efficiently blocked and, thus, the observed blueshifted \brg emission can be reproduced and yet produce sufficient 
spatial extension on the sky to also fit the AMBER visibilities. Models with $i > 60 \degr$ provide too much redshifted 
emission even if one of the hemispheres is completely blocked, while even for a polar density enhancement of 16 models 
with  $i < 30\degr$ do not provide enough spatial extension of the \brg emission region on the sky due to projection 
effects. If the inclination angle of the line-of-sight to \irc is less than $30\degr$, we estimate 
that a very large density contrast between pole and equator of $>30$ is required, which seems unlikely.

To summarize, from the 2-D modeling of our AMBER data we conclude that the wind of \irc has a significantly 
non-spherical geometry, and the viewing angle and the physical conditions must be such that a large fraction of 
the redshifted Br$\gamma$ emission is blocked from our view.

%%%%%%%%%%%%%%%%%%%%%%%%%%%%%%%%%%%%%%%%%%%%%%%%%%%%%%%%%
\section{Summary and conclusions}\label{sect_concl}
%%%%%%%%%%%%%%%%%%%%%%%%%%%%%%%%%%%%%%%%%%%%%%%%%%%%%%%%%

We presented the first VLTI/AMBER observations of the yellow hypergiant \irc in low-spectral
resolution (LR) mode covering the $J$, $H$, and $K$ bands and the first AMBER
observation of \irc around the \brg emission line in high-spectral resolution (HR) mode with
a spectral resolution of 12\,000 and projected baselines between 54 and 129~m. 

From the low-spectral resolution observations carried out with two linear AT arrays
(baselines ranging from 15 to 96~m), we were able to extract $H$- and $K$-band visibilities at 
spatial scales which probe the stellar continuum emission. Differential and closure phases could 
be derived from the LR data for two near-infrared bands (differential phases) and all three 
near-infrared bands (closure phases) covered by AMBER. Within the error bars, all LR phases turn 
out to be zero. We corrected all visibilities for AMBER's 
limited field-of-view (FOV = 250~mas for AT observations) to account for 
the truncation of \ircend's extended 
dust shell. From the visibilities we derived FWHM Gaussian diameters of \ircend's stellar component of 
$1.05\pm0.07$~mas and $0.98\pm0.10$, averaged over all spectral channels in the $H$- and $K$-bands, 
respectively. The stellar flux contribution in the case of an infinite FOV was found to be $0.55\pm0.02$ 
in the $H$ band and $0.60\pm0.02$ in the $K$ band.

On the other hand, we carried out high-spectral resolution AMBER observations of \irc centered
around the \brg emission line and combining three UTs. We found that the \brg emission is blueshifted 
with $v=-25~{\rm kms}$ with respect to the systemic velocity and that the \brg emitting-region of 
\ircend's dense stellar wind contributes roughly 50\% to the total flux at the peak wavelength of the 
nearly symmetric emission line. The equivalent width of the \brg line is $-6.7\AA$, in agreement with 
recent findings \cite[e.g.][]{dewit08}.
The HR visibilities of \irc show a strong decrease at the wavelength of the \brg-line emission, indicating
that the region in the wind of \irc where the \brg emission arises is fully resolved by our observations
and considerably more extended than the continuum-emitting region. In addition,
the HR AMBER observation of \irc revealed non-zero differential and closure phases with absolute values up 
to $\sim 30^\circ$ for several spectral channels covering the central and redshifted wing of the \brg-emission line. 
This finding indicates that the shape of the \brg line-emitting region is not point-symmetric. 

The HR data of \irc around the \brg emission line have been analyzed by simple geometrical models as well
as 1-D and 2-D gas radiative transfer models using the code CMFGEN. From simple geometrical models,
we found a size of the stellar continuum-emitting region comparable to our LR 
observations ($d_{\star} = 1.014\pm0.01$~mas).
Approximating the line-emitting region with an infinitely thin ring,
we found $d_{\rm ring} = 4.18^{+0.19}_{-0.08}$~mas, i.e.\ $d_{\rm ring} = 4.12 \times d_{\star}$.
The size of the line-emitting region is in basic agreement, although slightly larger than the value
3.3~mas found by \citet{dewit08} from AMBER observations in medium-spectral resolution mode
($\lambda/\Delta\lambda=1\,500$). The difference between the two results can
be explained by the differences in the details of the two modeling approaches and the different spectral
resolution of the data, but probably mainly reflects the overall calibration uncertainties.

To get a first idea of the deviation from sphericity, we fitted an elliptical Gaussian
ring to the 3 AMBER visibility data points at the central wavelength of the \brg emission. From this fit, we
find that the \brg line-emitting region is elongated, with major and minor FWHM ring diameters of $4.32\pm0.44$
and $3.85\pm0.37$~mas, and an elongation towards a position angle of $36\pm11$\degr, perfectly
aligned with the elongation axis of the outer reflection seen from HST images. Follow-up observations
with higher redundancy, a more robust calibration, and a better position angle coverage are indispensable 
to confirm this result.

Apart from simple geometrical models, we also analyzed the AMBER HR data of \irc around the \brg emission line
by comparison with 1-D and 2-D gas radiative transfer models obtained with the code CMFGEN. 
We found that spherical models definitely fail to reproduce the observed blueshift of the \brg emission
while they can well reproduce the amount of line emission. Spherical wind models also fail to reproduce, 
at the same time, the amount of line emission and the wavelength dependence of the visibility across the 
emission line and, hence, the extension of the \brg line-emitting region.
On the other hand, we found that 2-D models with a prolate wind structure can simultaneously reproduce the amount 
of \brg emission, the wavelength dependence of the visibility across the line and, in addition, the 
observed blueshift of the \brg emission, but only if a substantial fraction of the redshifted 
\brg emission is blocked. A model without any diminishing of the redshifted emission cannot explain the
blueshift of the emission while {\it simultaneously} reproducing the amount of line emission and the 
spatial extension of the line-emitting region. In our best-fitting model, the outflow exhibits a strong 
density enhancement towards the poles with a pole-to-equator density ratio of 16, and the outflow is oriented 
towards a PA of 60\degr and inclined by 45\degr.
Since the uncertainty of the position angle of the best-fitting model is approximately 20\degr, we find that the
orientation of the symmetry axis of our radiative transfer model is in rough agreement with the orientation
of the elongation derived from our simple elliptical ring model ($\phi=36\pm 11^\circ$, see Sect.~\ref{sect_res_hr_mod}).
Nevertheless, it should be noted that both results might be biased towards the derived values due to the 
sparse position angle coverage of our measurements.

Interestingly, our best-fitting model gives a temperature of 1500~K at a radial distance of 100~mas 
from the central star, assuming a distance of 5 kpc for direct comparison with the \citet{bloecker99} results. 
In their best model, based on speckle-interferometric observations from 1998,
the inner dust shell boundary with a dust temperature of 1000~K is located at a radial distance of $\sim70$~mas.
Thus, our gas-radiative transfer model nicely fits the previously developed dust-shell model, especially if
we believe that we indeed see an outward shift of the inner dust shell boundary which is driven by the noticeable
increase of the stellar effective temperature within the last decades.

The signature of the differential- and closure-phase signal of \irc detected 
with AMBER is totally different from what would be expected, for instance, for a rotating circumstellar 
disk \citep[e.g.][]{meilland07}. On the other hand, we find that the wavelength dependence of both differential 
and closure phases shows some qualitative similarities to measurements obtained for the Luminous Blue Variable 
$\eta$~Carinae \citep{weigelt07}. In the interpretation of
the $\eta$~Car AMBER data, the phase signature was associated with enhanced mass loss of the rapidly
rotating, massive primary star in polar direction, and the observer is basically looking towards the blueshifted 
part of the outflow. The qualitative similiarty of the phase signals between \irc and \brg motivates the assumption 
that also in the case of \irc rapid rotation drives a non-spherical and polar enhanced mass loss as inferred from 
theoretical considerations \citep[e.g.][]{owocki98}. This would be in line with the fact that \irc is a massive star 
in a highly evolved evolutionary stage and would also qualitatively fit to the observed blueshift of the \brg line.
However, the major difference between the differential and closure phases of $\eta$~Car and \irc is that
in \irc the phase signal in stronger in the redshifted wing of the \brg emission and not in the blueshifted wing as
in the case of $\eta$~Car. This makes the modeling of the \irc data more puzzling and probably indicates that the
complexity of the overall structure of the innermost wind region \irc goes well beyond the capabilities
of a 2-D modeling as carried out in this paper. 

%For instance, since our modeling is restricted to an outflow geometry, the fact that we were not able to reproduce 
%all AMBER observables simultaneously, in particular the phases, might indicate that \irc exhibits an inflow rather
%than an outflow, or maybe matter infall and outflow are present at the same time \citep{humphreys02}, for example 
%in terms of an equatorial infall and a polar outflow. Such an equatorial infall could lead to the formation of, e.g., 
%an optically thick disk which might be able to explain the blocking of the
%redshifted wing as well as the strong phase signal in the redshifted part of the \brg emission. On the other hand,
%the fact that we have to artificially block a substantial fraction of the redshifted line emission in our model could
%also indicate that the mass outflow from \irc is indeed highly asymmetric. We can only speculate on the reason for such
%a strong asymmetry in the outflow but, in principle, such asymmetries could be connected to large-scale 
%asymmetries in the surface structure of the central star (e.g., hot spots) which lead to prefered directions 
%of mass ejection \citep{humphreys02}, or they could be triggered by a yet undetected binary companion.

Nevertheless, our modeling indicates that a substantial fraction of the redshifted line emission does not reach the 
observer. Since our modeling is restricted to an axisymmetric outflow geometry, the fact that we were not able to 
reproduce all AMBER observables simultaneously (in particular the phases) might indicate the following:
\begin{itemize}
\item[$\bullet$]
The mass outflow from \irc is indeed intrinsically highly asymmetric. In this case, one of the hemispheres 
(which is pointing towards us and, thus, emitting blueshifted \brg) has a much higher mass-loss rate than the other 
hemisphere (responsible for the redshifted light). We can only speculate on the reason for such
a strong asymmetry in the outflow but, in principle, such asymmetries could be connected to large-scale asymmetries in 
the surface structure of the central star (e.g., hot spots) which lead to preferred directions of mass ejection 
\citep{humphreys02}. In this context, it is interesting to note that recent CO measurements \citep{castro07,TRUNG09}
detected a flux deficit in the CO shells in the south-western direction. Thus, \irc seems to exhibit an
asymmetry in the same direction as found in our study also on much larger spatial scales, indicating that this asymmetry
might be of a long-term nature.
\item[$\bullet$]
The wind of \irc is axisymmetric, but the redshifted light is blocked by an optically-thick disk which 
would also explain the strong phase signal in the redshifted part of the \brg emission. Interestingly, such a disk has 
to be located very close to the star ($\sim 5$ mas)  in order to block the \brg emission, and would have to be formed 
of mainly neutral+molecular gas in order to effectively absorb the redshifted light, since the presumed dust sublimation 
radius is located much farther out, at scales of $\sim 100$ mas. An yet undetected binary companion could also cause 
the presence of an equatorial disk.
\item[$\bullet$]
\irc exhibits an inflow rather than an outflow, or maybe matter infall and outflow are present at the same 
time \citep{humphreys02}, for example in terms of an equatorial infall and a polar outflow. Such an equatorial infall 
could lead to the formation of an optically thick disk.
\end{itemize}
Of course, the above scenarios are not mutually exclusive, and both could contribute to explain the peculiar properties 
seen in \irc.

The study presented here has illustrated AMBER's great potential to probe the innermost circumstellar
gas environment of evolved stars with spectral resolutions up to 12\,000. New observations of \irc 
covering more baselines and position angles, which are foreseen for forthcoming observing periods, 
will undoubtably provide a much clearer picture of the geometry of \ircend's inner wind zone. For instance,
new observations will help to support or disprove the model according to which \irc is exhibiting a 
prolate wind structure with a substantial shielding of the redshifted emission.
Thus, further AMBER observations probing the inner wind zone of \irc
will help to answer long-standing questions on the physical conditions and the mass-loss processes of
this outstanding, rapidly evolving object and shed more light on the question of the mass-loss 
conditions in highly evolved massive stars.

%%%%%%%%%%%%%%%%%%%%%%%%%%%%%%%%%%%%%%%%%%%%%%%%%%%%%%%%%%%%%%%%%%%%%%%%%%%%%%%%%%%%%%%%%%%%%%%%%%%%%
\begin{acknowledgements}
%%%%%%%%%%%%%%%%%%%%%%%%%%%%%%%%%%%%%%%%%%%%%%%%%%%%%%%%%%%%%%%%%%%%%%%%%%%%%%%%%%%%%%%%%%%%%%%%%%%%%

We thank the ESO VLTI team on Paranal and in Garching for carrying out the AMBER observations
presented in this paper. The data presented here were reduced using the
publicly available data reduction software package {\it amdlib} kindly provided by 
the Jean-Marie Mariotti Center (http://www.jmmc.fr/data\_processing\_amber.htm).
We warmly thank John Hillier and Joe Busche for making CMFGEN and the \citet{bh05} codes available, 
and for continuous support with the codes.
The high spectral resolution telluric spectra used in this work for spectral calibration
of the AMBER data were created from data that was kindly made available by the NSO/Kitt Peak 
Observatory. Finally, we thank the anonymous referee for valuable comments which 
helped to considerably improve the manuscript.

\end{acknowledgements}
%%%%%%%%%%%%%%%%%%%%%%%%%%%%%%%%%%%%%%%%%%%%%%%%%%%%%%%%%%%%%%%%%%%%%%%%%%%%%%%%%%%%%%%%%%%%%%%%%%%%

%%%%%%%%%%%%%%%%%%%%%%%%%%%%%%%%%%%%%%%%%%%%%%%%%%%%%%%%%%%%%%%%%%%%%%%%%%%%%%%%%%%%%%%%%%%%%%%%%%%%
\bibliographystyle{aa}
\bibliography{11723}

\end{document}